\documentclass[lettersize,journal]{IEEEtran}
\usepackage{amsmath,amsfonts}
\usepackage{algorithmic}
\usepackage{array}
\usepackage[caption=false,font=normalsize,labelfont=sf,textfont=sf]{subfig}
\usepackage{textcomp}
\usepackage{stfloats}
\usepackage{url}
\usepackage{verbatim}
\usepackage{graphicx}
\def\BibTeX{{\rm B\kern-.05em{\sc i\kern-.025em b}\kern-.08em
T\kern-.1667em\lower.7ex\hbox{E}\kern-.125emX}}
\usepackage{balance}

\usepackage{threeparttable}
\usepackage{tabularx}
\usepackage{color}
\usepackage{booktabs}
\usepackage{verbatim}
\usepackage{amsopn}
\usepackage{amsmath}
\usepackage{multirow}
\usepackage{colortbl}
\usepackage{makecell}
\usepackage{colortbl}
\definecolor{mygray}{gray}{.9}
\definecolor{orange}{rgb}{1,0.5,0}
\usepackage{colortbl}
\usepackage{soul}
\usepackage{amsfonts,amssymb}
\usepackage{bbm}
\usepackage{bm}
\usepackage{orcidlink}

\usepackage{hyperref}
\hypersetup{
colorlinks=true,
urlcolor=magenta,
pdftitle={Overleaf Example},
pdfpagemode=FullScreen,
}

\usepackage{xspace}

\makeatletter
\DeclareRobustCommand\onedot{\futurelet\@let@token\@onedot}
\def\@onedot{\ifx\@let@token.\else.\null\fi\xspace}

\def\eg{\emph{e.g}\onedot} 
\def\ie{\emph{i.e}\onedot} 
 
\def\etc{\emph{etc}\onedot} 
 
\def\etal{\emph{et al}\onedot}
\makeatother
\usepackage{epstopdf}

\begin{document}
\title{Exploring Sparse Spatial Relation in Graph Inference for Text-Based VQA}
\author{Sheng Zhou$^{\orcidlink{0009-0007-4215-5464}}$, Dan Guo$^{\orcidlink{0000-0003-2594-254X}}$, Jia Li$^{\orcidlink{0000-0001-9446-249X}}$, Xun Yang$^{\orcidlink{0000-0003-0201-1638}}$, and Meng Wang$^{*}$$^{\orcidlink{0000-0002-3094-7735}}$, \emph{IEEE Fellow} 
\thanks{
This work was supported in part by the National Key R\&D Program of China (2022YFB4500600), the National Natural Science Foundation of China (Grant 62020106007, Grant 62272144, Grant U20A20183, Grant 72188101, Grant 62272435, Grant U22A2094, and Grant 62202139), and the Major Project of Anhui Province (202203a05020011). \emph{(Corresponding authors: Dan Guo, Xun Yang, Meng Wang.)}

Sheng Zhou, Dan Guo, Jia Li, and Meng Wang are with the School of Computer Science and Information Engineering, Hefei University of Technology, Hefei, 230601, China (email: hzgn97@gmail.com; guodan@hfut.edu.cn; jiali@hfut.edu.cn; eric.mengwang@gmail.com). 

Xun Yang is with the School of Information Science and Technology, University of Science and Technology of China, Hefei, 230026, China. (email: xyang21@ustc.edu.cn)
}}

\markboth{IEEE TRANSACTIONS ON IMAGE PROCESSING,~Vol.~32, 2023}%
{How to Use the IEEEtran \LaTeX \ Templates}

\maketitle

\begin{abstract}
Text-based visual question answering (TextVQA) faces the significant challenge of {avoiding redundant relational inference}. To be specific, a large number of detected objects and optical character recognition (OCR) tokens result in rich visual relationships. Existing works take all visual relationships into account for answer prediction. However, there are three observations:  
(1) a single subject in the images can be easily detected as multiple objects with distinct bounding boxes (considered repetitive objects). The associations between these repetitive objects are superfluous for answer reasoning; 
(2) two spatially distant OCR tokens detected in the image {frequently have weak semantic dependencies for answer reasoning}; and (3) the co-existence of nearby objects and tokens may be indicative of important visual cues for predicting answers. Rather than utilizing all of them for answer prediction, we make an effort to identify the most important connections or eliminate redundant ones.  
We propose a sparse spatial graph network (SSGN) that introduces a spatially aware relation pruning technique to this task. As spatial factors for relation measurement, we employ spatial distance, geometric dimension, overlap area, and DIoU for spatially aware pruning. We consider three visual relationships for graph learning: object-object, OCR-OCR tokens, and object-OCR token relationships. {SSGN is a progressive graph learning architecture that verifies the pivotal relations in the correlated object-token sparse graph, and then in the respective object-based sparse graph and token-based sparse graph.}
Experiment results on TextVQA and ST-VQA datasets demonstrate that SSGN achieves promising performances. And some visualization results further demonstrate the interpretability of our method.
\end{abstract}

\begin{IEEEkeywords}
Visual question answering, text-based visual question answering, graph inference, spatial relation, and relation learning.
\end{IEEEkeywords}

\section{Introduction}
\IEEEPARstart{S}{cene} text expresses rich information in human activities, such as numeric symbols \cite{gao2020multi}, advertisement slogans \cite{singh2019towards}, traffic signs \cite{biten2019scene}, and price tags in shops \cite{biten2019scene}. Text-based visual question answering (TextVQA) becomes an emerging hot topic in the field of vision and language. The TextVQA models have a wide range of applications, such as visual impairment assistance, online education, online shopping \cite{Zhu2021SimpleIN}, driving assistance \cite{Zhu2021SimpleIN}, \emph{etc}.

\begin{figure}[tp]
\centering
\includegraphics[width=\columnwidth]{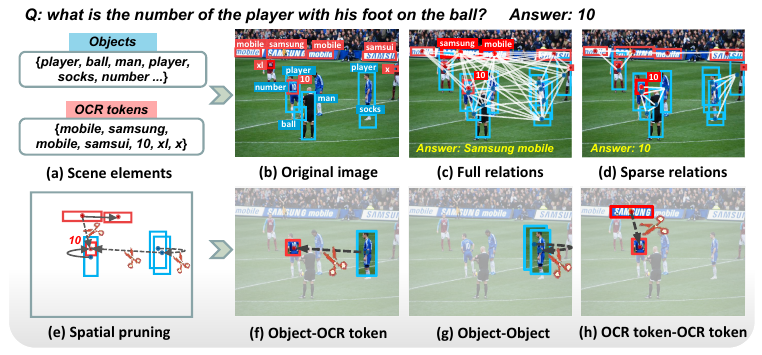}
\caption{Full (dense but redundant) relations vs. sparse relations in TextVQA task. The OCR system and the object detector have their own characteristics. The OCR system performs well at scene text recognition (\eg, numbers, symbols, characters), whereas the object detector excels at identifying visual subjects (\eg, people, animals, substances). We have to understand the scene texts {or objects} queried by the question both well. We take all the detected objects and OCR tokens in the image as visual entity nodes. Our purpose is to build an effective sparse spatial graph on the basis of spatial-aware relation pruning for answer prediction. }
\label{Figure 1}
\vspace{-0.3cm}
\end{figure}

With the advancement of artificial intelligence technology, many multimodal analysis models have been rapidly developed with visual understanding capabilities \cite{wang2022progressive, guo2018hierarchical, guo2019dadnet, guo2019dense}, such as the tasks of image captioning \cite{Ji2020SpatioTemporalMA}, {visual grounding \cite{Liao2022ProgressiveLV}}, and visual question answering (VQA) \cite{Guo2021ReAttentionFV}. The most relevant research to our work is the VQA task. General VQA models \cite{Gu2021GraphBasedMN} have powerful reasoning capabilities to answer object-based visual questions regarding people, scenes, motifs, and even plot comprehension. However, the TextVQA models are dedicated to discovering the scene texts and utilizing them to answer the text-based visual questions, where scene texts may involve small, fuzzy, and illegible text fonts.  To make up this research gap, Singh \etal \cite{singh2019towards} have released a novel TextVQA task and a new TextVQA dataset referring to both object-related and text-related visual questions. Meanwhile, Bitten \etal \cite{biten2019scene} have released another new dataset, ST-VQA, which could only answer questions using the scene text in the image. {By comparison, scene text is extremely critical in the TextVQA task, for example, all questions in the ST-VQA \cite{biten2019scene} dataset are related to scene text.}

Many efforts have been made to solve this task. Just in time, optical character recognition (OCR) tasks \cite{Baek2019WhatIW, Fang2021ReadLH} have made significant progress in the field of computer vision. Under the research background, some researchers \cite{singh2019towards, biten2019scene, hu2020iterative, gao2020multi, kant2020spatially, liu2020cascade} apply this technique to existing VQA models, enabling the models with the ability to read scene text accurately. For example, LoRRA~\cite{singh2019towards} is the first backbone for TextVQA which extends the VQA model Pythia~\cite{jiang2018pythia} with a new OCR attention branch; the model is enabled to select answer words from a predefined vocabulary set of objects and an online set of OCR tokens, where the vocabulary set and the OCR set are collected from the training set and each image itself respectively. 
Besides, based on the success of Transformer \cite{vaswani2017attention} and BERT \cite{Devlin2019BERTPO}, M4C \cite{hu2020iterative} implements a multi-modal transformer with a multi-step response which serves as another backbone and is widely-used for existing methods. Based on above two backbones, some graph models have been introduced into TextVQA task because of its outstanding relation reasoning ability, \eg, MM-GNN \cite{gao2020multi}, SA-M4C \cite{kant2020spatially}, and CRN \cite{liu2020cascade}. For a better exploration of visual relations, we adopt the graph structure in this work. 

Almost the aforementioned works share a common feature in that the full relations among the visual entities including objects and OCR tokens are exploited to predict answers. Differently, we argue that excessive redundant relations exist among these visual entities, which may mislead the prediction of answers. As shown in Fig.~\ref{Figure 1} (c), driven by the question of ``what is the number of the player with his foot on the ball?", under the full relations, ``samsung mobile" is predicted as the answer due to the dense relations with the other visual entities, while the correct answer is ``10". 

This study is proposed to solve the relation redundancy problem for the TextVQA task. To be specific, we propose a sparse graph network inspired by three observations: (1) multiple bounding boxes covering a same visual entity are interpreted as different objects by the object detectors accompanied with redundant connections, {\eg, the  rightmost player is labeled with three bounding boxes in Fig. \ref{Figure 1} (c)}; (2) distant OCR tokens often have weak or no semantic dependencies, {\eg, the two OCR tokens ``samsung'' and ``10'' in Fig. \ref{Figure 1} (c)}; (3) association between remote objects and tokens are useless, and the visual regions detected with both object and OCR token labels are informative for answer prediction, {\eg, ``10'' is detected as ``number" (object label) and number ``10'' (OCR token label) in Fig. \ref{Figure 1} (c).} In short, the idea originates from the native perspective of spatial relation. As shown in Fig. \ref{Figure 1} (e), the spatial factors in our work refer to Distance-IoU \cite{zheng2020distance}, relative distance, geometric size, and overlap area. We work to use the spatial factors to eliminate the redundant relations between remote object and OCR token in Fig. \ref{Figure 1} (f), much-overlapping objects in Fig. \ref{Figure 1} (g), and  distant OCR tokens in Fig. \ref{Figure 1} (h).  

\begin{figure*}[t]
\centering
\includegraphics[width=\textwidth]{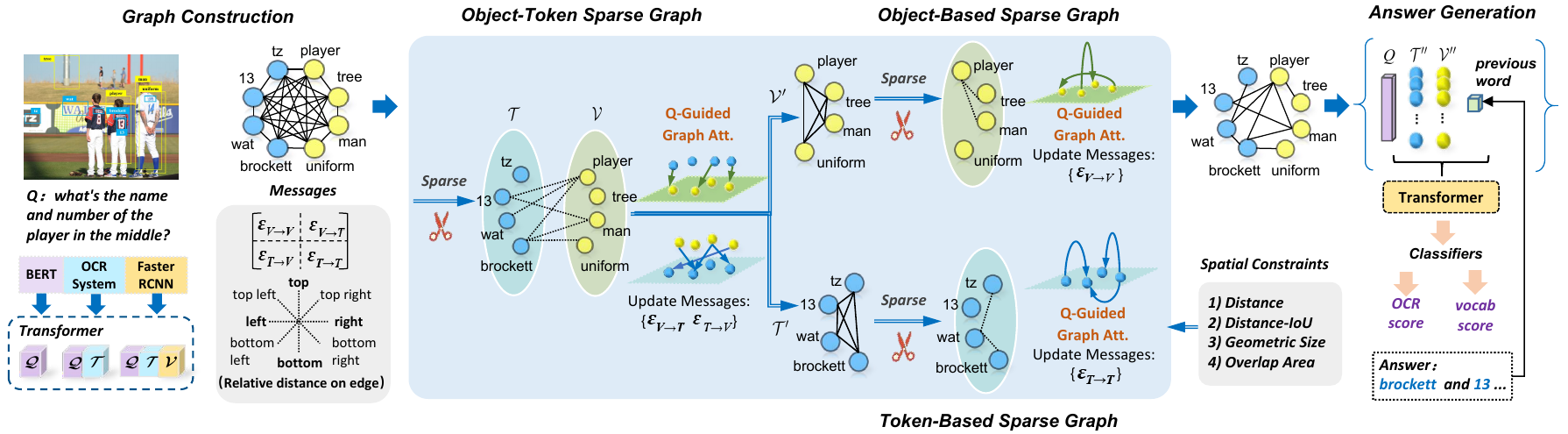}
\caption{The overall framework of Sparse Spatial Graph Network (SSGN). Given an image and a question, we prepare the features of the question  ($\mathcal{Q}$), OCR tokens ($\mathcal{T}$), and objects ($\mathcal{V}$). After that, we construct graphs with $\mathcal{T}$ and $\mathcal{V}$ and build edges with the condition of spatial relation. The relations in the graph are defined as $\mathcal{E}$ =
$({\mathcal{E_{V \to V}}}$, ${\mathcal{E_{T \to T}}}$, ${\mathcal{E_{V \to T}}}$, ${\mathcal{E_{T \to V}}})$. In this work, graph inference is used to update OCR token and object features under the guidance of the question $\mathcal{Q}$. We propose a novel progressive graph learning scheme with partial context update---the message (${\mathcal{E_{V \to T}}}$, ${\mathcal{E_{T \to V}}}$) updates first and then  ${\mathcal{E_{V \to V}}}$ and ${\mathcal{E_{T \to T}}}$ update in parallel. Finally, we use an answer generation module to iteratively predict answers.}
\vspace{-0.3cm}
\label{Figure 2}
\end{figure*}

To achieve effective relation reasoning, we propose a {\textbf{S}parse \textbf{S}patial \textbf{G}raph \textbf{N}etwork \textbf{(SSGN)}}, which cuts off useless or negative relations (edges in the graph) between numerous visual entities (object and OCR token nodes) to suppress the redundant message passing. As shown in Fig.~\ref{Figure 2}, SSGN is a hierarchical graph learning architecture by excluding redundant relations (we deem these as noises) in three sparse sub-graphs. (1) {{\textbf{O}bject-\textbf{T}oken  \textbf{S}parse \textbf{G}raph, \textbf{OTSG}}} first removes redundant relations to learn object-OCR token correlation verification, (2) {\textbf{O}bject-based \textbf{S}parse \textbf{G}raph, \textbf{OSG}}, and (3) {\textbf{T}oken-based \textbf{S}parse \textbf{G}raph, \textbf{TSG}} then verify the pivotal relations of \textbf{OTSG} in each visual entity space (object or OCR token). All the above three sub-graphs are implemented under the guidance of question. Based on this framework, SSGN progressively updates object and OCR token features by the graph learning of OTSG, OSG and TSG for answer prediction. Experiments conducted on TextVQA and ST-VQA datasets show that SSGN outperforms other comparative methods in the TextVQA task. Extensive ablation studies and visualization results demonstrate the validity, robustness, and interpretability of our method.

{The main contributions are summarized as follows.}
\begin{itemize}
\item \emph{For task}, {we perform a sparse spatial graph learning by introducing a spatial-aware relation pruning method into the TextVQA task, which shields redundant relations among numerous visual entities in complex scenes.}

\item \emph{About sparsity}, {we propose a graph preprocessing method which utilizes spatial coordinates of visual entities to build sparse relations (edges). In this work, we primarily emphasize the spatial-constrained relation pruning to trim the redundant relations. Besides, during graph learning, the pruned relations pass messages between visual entities under the guidance of question semantics.}

\item \emph{About spatiality}, {we consider Distance-IoU (DIoU), spatial distance, geometric size, and overlap area as the primary spatial factors in relation modeling.
The motivation is that spatially distant visual entities in natural images often have less or even no relations, and geometrically similar boundary boxes with large overlaps may exist as superfluous visual entities.}

\item \emph{About graph Inference}, {we propose a hierarchical graph learning solution with partial edges to achieve node update. We first perform double-sided correlation verification (object-OCR token) in the OTSG, and then further verify the pivotal relations in respective OSG (object) and TSG (OCR token).}
\end{itemize}

The rest of the paper is organized as follows. We provide an overview of related work in Sec.~\ref{subsec:Related Work} and detail the proposed SSGN method in Sec.~\ref{subsec:Method}. Extensive experiments including quantitative comparison with state-of-the-art methods, ablation studies, and visualization analysis are presented in Sec.~\ref{subsec:Experiment}, followed by a brief summary of this work in Sec.~\ref{subsec:Conclusion}. 

\vspace{-0.1cm}

\section{Related Work}
\label{subsec:Related Work}
\subsection{Text-Based Visual Question Answering}
With the development of visual question answering tasks, some models \cite{Kafle2018DVQAUD, Hudson2019GQAAN} have been proposed to improve the reasoning ability.
Recently, TextVQA task \cite{singh2019towards} has become a new research focus aimed at answering questions about scene text comprehension. To promote this study, there are two benchmark datasets, TextVQA \cite{singh2019towards} and ST-VQA \cite{biten2019scene}, and two backbones, LoRRA \cite{singh2019towards} and M4C \cite{hu2020iterative}. LoRRA \cite{singh2019towards} uses an off-the-shelf OCR system \cite{borisyuk2018rosetta} to detect multiple OCR tokens in the image and extends a previous VQA model \cite{jiang2018pythia} to select a single OCR token as the answer. The methodological bottleneck of LoRRA is the inability to generate multiple words in an answer. To solve this problem, M4C \cite{hu2020iterative} introduces a pointer-augmented multi-step decoder based on a transformer architecture to generate multiple words in the answer. Based on the above two backbones, existing methods could mainly be divided into four categories: (1) \textbf{{Attention-based model.}}  SSBaseline \cite{Zhu2021SimpleIN} introduces an attention mechanism into M4C \cite{hu2020iterative} to achieve feature aggregation, which greatly reduces the computation and obtains a good performance; (2) \textbf{{Transformer-based model.}} TAP \cite{yang2021tap} introduces large-scale data and pre-training techniques in a transformer architecture to improve performance. PAT \cite{zhang2021position} adopts a position-augmented transformer with entity-aligned mesh  for comprehensively capturing position relations of visual entities; (3) \textbf{{Graph-based model.}} There are some graph networks focus on relation inference between objects and OCR tokens, including multi-modal graph (MM-GNN) \cite{gao2020multi}, spatial-aware graph (SA-M4C) \cite{kant2020spatially}, role-aware graph (SMA) \cite{gao2020structured}, and cascade reasoning network (CRN) \cite{liu2020cascade}; (4) \textbf{{Representation learning model.}}  BOV \cite{zeng2021beyond} proposes a visually enhanced scene text embedding and an object-oriented embedding to enhance the feature representation and improve the reasoning ability.

Previous works develop feature fusion \cite{zeng2021beyond}, feature alignment \cite{zhang2021position, yang2021tap}, feature attention \cite{Zhu2021SimpleIN}, and relation learning \cite{gao2020multi, kant2020spatially, gao2020structured, liu2020cascade} to address the TextVQA task. Among them, graph model in TextVQA task shows outstanding advantage. {Most graph models take full relations between or within objects and OCR tokens into account. Differently, we devote
to removing redundant relations between object-object, OCR-OCR tokens, and object-OCR token to achieve effective answer reasoning.}


\subsection{Graph Inference Technique for Visual Reasoning}
Plenty of works demonstrate that graph neural network (GNN) has a strong ability for relation reasoning \cite{li2019relation,kim2020hypergraph}. Both intra-modal and inter-modal graph models are applied for addressing the visual relation reasoning issue in various VQA and visual dialog tasks \cite{Zhu2020MuckoMC,chen2021gog,huang2020aligned}. 
In general VQA task, Huang \etal \cite{huang2020aligned} propose a novel dual-channel graph convolutional network (DC-GCN) to capture the visual relations between objects and the syntactic relations between question words separately. 
For Fact-based VQA \cite{Zhu2020MuckoMC}, Zhu \etal first construct three intra-modal graphs to separately explore visual, semantic, and knowledge clues, and then aggregate them through cross-modal graph convolutions. As for visual dialog, Chen \etal design a graph-over-graph network named GoG \cite{chen2021gog} with three graphs for exploring co-reference relation among dialog history, dependency relation between question words, and spatial relation between visual objects.  These relation-aware graphs are proposed to exploit consistent contexts from vision (image) and text (question) for visual reasoning. Similar to previous GNN works \cite{Zhu2020MuckoMC,chen2021gog,huang2020aligned}, we utilize the heterogeneous graph structures (intra- and inter-modality graphs) for relation inference. Besides, these previous GNN work \cite{Zhu2020MuckoMC,chen2021gog,huang2020aligned} usually explores the fully-connected graph learning. The difference of this work is it extends a novel hierarchical graph learning scheme with sparse relations (edges) for achieving reasonable relation reasoning.

\subsection{Exploitation of Spatial Relation in Vision}
For visual reasoning, we insist that the exploitation of spatial relation can facilitate answer inference. We investigate the related work and find that the exploitation of visual-spatial relation is indeed beneficial for quite a few vision-related tasks \cite{chen2020monopair,wang2020multimodal,inayoshi2020bounding,li2019relation,yao2018exploring}. For example, to fully understand the visual scene in the VQA task, ReGAT \cite{li2019relation} utilizes a graph network to model the spatial relations between objects by measuring relative angles and overlapping areas of two objects in the image.
For the TextCaps task \cite{sidorov2020textcaps}, Wang \etal \cite{wang2020multimodal} strengthen the semantic correlation between two OCR tokens from both horizontal and vertical position dimensions to generate the captions, which is motivated by the spatial orientation relation of OCR token pairs. For object detection,  Chen \etal \cite{chen2020monopair} present a new object detector to locate the required partially-occluded object in the image, which modifies the bounding box position by measuring the spatial distance of  the occluded object and its neighbors. As for the visual relation detection task, Inayoshi \etal \cite{inayoshi2020bounding} propose a boundary-box channel-wise fusion method, which introduces object position and overlapping area into the image features for better identifying the relation between objects.

There is merely single object-object relation or OCR-OCR tokens relation discussed in the above tasks. We leverage the spatial relations between and within the two visual entities (\ie, object, and OCR token) for TextVQA. To explore the spatial relation for visual reasoning, we perform a multi-view spatial measurement by the spatial facts of DIoU, relative distance, geometric size, and overlapping area, which are well-designed for keeping the characteristic of the object and OCR token and refine the message passing between object-OCR token, object-object, and OCR-OCR tokens in the graph structure.

\section{Method: Sparse Spatial Graph Network}
\label{subsec:Method}
In this work, we devote to addressing the relation reasoning between the detected objects and OCR tokens for the TextVQA task. We propose a Sparse Spatial Graph Network (named SSGN) for relation inference. As shown in Fig. \ref{Figure 2}, the question answering process involves three steps: (1) obtaining features of objects, OCR tokens, and the question, and then building a spatial-aware graph network with objects and OCR tokens (detailed in Sec.~\ref{subsec:Graph Construction}), (2) performing spatial-aware relation pruning and implementing a hierarchical sparse spatial graph learning, where the graph involves the object-object, OCR-OCR tokens, and object-OCR token relations (the core part of our method introduced in Sec.~\ref{subsec:Sparse Spatial Graph Network}), and (3) updating the features of object and OCR token nodes in the graph and feeding them into an iterative answer decoder for answer prediction (detailed in Sec.~\ref{subsec:Answer Generation Module}). 


\subsection{Preliminary}
\label{subsec:Graph Construction}
\subsubsection{\textbf{Feature Preparation}}
\label{subsec:Feature Preparation}
For an image $I$, we extract the initial object features by pre-trained Faster R-CNN \cite{Ren2015FasterRT}. The initial OCR token features are extracted by OCR systems, such as Rosetta-en \cite{borisyuk2018rosetta}, SBD-Trans \cite{Liu2019OmnidirectionalST}, Google-OCR \footnote[1]{Google-OCR API: \href{https://cloud.google.com/products/ai/}{https://cloud.google.com/products/ai/}\label{fontnote1}}, and Microsoft-OCR \footnote[2]{Microsoft-OCR API: \href{https://azure.microsoft.com/en-us/services/cognitive-services/computer-vision/}{https://azure.microsoft.com/en-us/services/cognitive-services/computer-vision/}\label{fontnote2}}. Besides, we obtain a initial question feature by a fine-tuned {three-layer BERT-BASE \cite{Devlin2019BERTPO} where the number of hidden layer in BertEncoder is set to 3}. Following the previous work \cite{liu2020cascade}, we input the initial features of the objects, OCR tokens, and the question into a transformer-based encoder architecture and update the features to $\mathcal{Q}$$=$$\{q_i\}_{i=1}^K$, $\mathcal{V}$$=$$\{v_i\}_{i=1}^N$, and $\mathcal{T}$$=$$\{t_i\}_{i=1}^M$, $v_i, t_i, q_i \in\mathbb{R}^{d}$, where $N$, $M$, and $K$ is the number of objects, OCR tokens, question words, respectively.

\subsubsection{\textbf{Spatial-Aware Graph}}
To acquire the relations of object-object, OCR-OCR tokens, and object-OCR token, we construct a spatial-aware graph $\mathcal{G}$$=$$\left\{\mathcal{N},\mathcal{E}\right\}$, where $\mathcal{N}$ is a node set that includes all the objects and OCR tokens, $\mathcal{N}$ = $\mathcal{V} \cup \mathcal{T}$, and $\mathcal{E}$ is a directed edge set denoted as $\mathcal{E}$ =
$({\mathcal{E_{V \to V}}}$, ${\mathcal{E_{T \to T}}}$, ${\mathcal{E_{V \to T}}}$, ${\mathcal{E_{T \to V}}})$. Among ${\mathcal E}$, ${\mathcal{E_{V \to V}}}$ and ${\mathcal{E_{T \to T}}}$ are two subsets that describe object-object and OCR-OCR tokens relations respectively; ${\mathcal{E_{V \to T}}}$ and ${\mathcal{E_{T \to V}}}$ are two edge subsets that contain the object-OCR edges with different message passing directions, \ie, from object to OCR token, and from OCR token to object. To make full use of the spatial information in the image, we use spatial coordinates of visual entities to model their explicit relations (edges in the graph). To be specific, we consider the relative distance and height-width ratio of the boundary boxes of two visual entities in the image. Taking $\mathcal{E}_{i \to j}$ as an example, we take node $n_j$ as the reference node and measure the distance of node $n_i$ to it. {The edge feature vector $\mathcal{E}_{i \to j} \in \mathbb{R}^{d_e}$ is encoded below, where $d_e = 5$ }.
\begin{equation}{
\mathcal{E}_{i \to j}\!=\!\left[ {\frac{{x_i^{tl}\!-\!x_j^c}}{{{w_j}}}\!,\!\frac{{y_i^{tl}\!-\! y_j^c}}{{{h_j}}}\!,\!\frac{{x_i^{br}\!-\!x_j^c}}{{{w_j}}}\!,\!\frac{{y_i^{br}\!-\!y_j^c}}{{{h_j}}}\!,\!\frac{{{w_i}\!*\!{h_i}}}{{{w_j}\!*\!{h_j}}}} \right],
}
\label{eq:eq5}
\end{equation}
where $\left[{x_i^{tl},y_i^{tl}], [x_i^{br},y_i^{br}} \right]$ denote the top-left and bottom-right coordinates of the bounding box of node $n_i$,  $[x_i^c,y_i^c]$ and $[{w_i},{h_i}]$ denote the center coordinate, width and height of bounding box of node $n_i$.

\subsection{Sparse Spatial Graph}
\label{subsec:Sparse Spatial Graph Network}

\begin{figure}[tp]
\centering
\includegraphics[width=\columnwidth]{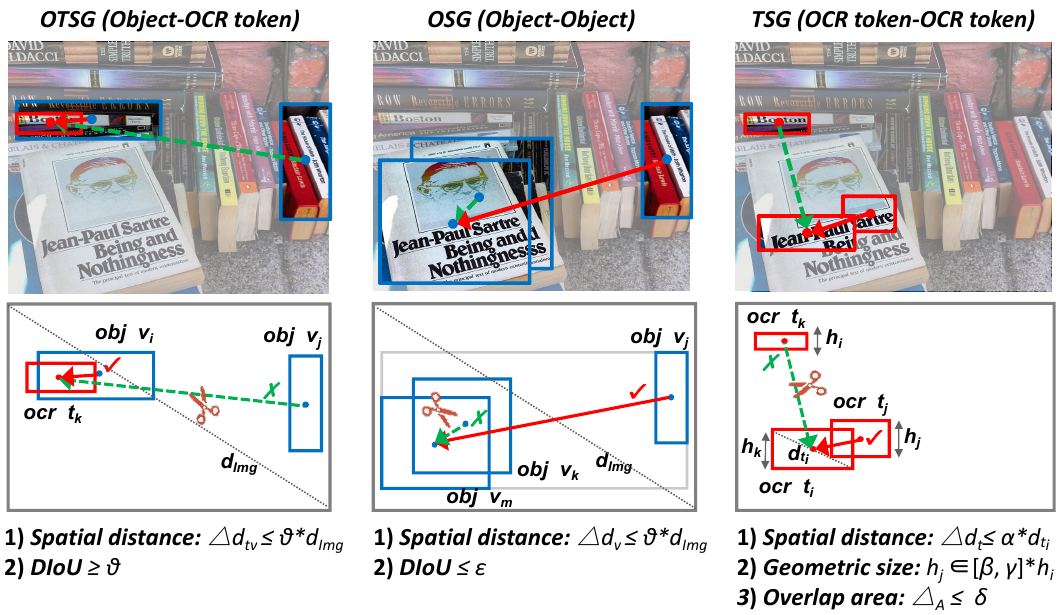}
\caption{Relation pruning based on spatial factors. We expect to establish the relations by co-occurring nearby object and token pair in OTSG and remove the redundancy of overlapping objects in OSG and distant tokens in TSG.}
\label{Figure 3}
\vspace{-0.3cm}
\end{figure}

Each image contains a plethora of visual entities including objects and OCR tokens. The TextVQA task is characterized by reasoning out one or a few specified scene texts or objects in a local region of an image to answer the question. Driven by the question, there are many unnecessary visual relations in reasoning process that may cause interference. {Motivated by the visual spatial relations in the natural image scene}, as shown in Fig.~\ref{Figure 3}, we seek to remove the redundant relations of much-overlapping objects and distant tokens, and enhance the relations by co-occurring nearby object and token pairs. 

As shown in Fig. \ref{Figure 2}, we perform a hierarchical graph inference. The whole process is implemented under the question guidance. About the relation pruning,  {Object-Token Sparse Graph} (Sec. \ref{subsec:Cross-modal sparse graph}) first performs the correlation of nearby objects and OCR tokens to filter out useless object-OCR relations. Next, we conduct the {Object-Based Sparse Graph} (Sec. \ref{subsec:Object-based Sparse Graph}) and {Token-Based Sparse Graph} (Sec. \ref{subsec:Token-based Sparse Graph}) in each entity space (\emph{i.e.}, object or OCR token) in parallel, which further refines the correlated relations for answer prediction. 

\subsubsection{\textbf{Object-Token Sparse Graph (OTSG)}}
\label{subsec:Cross-modal sparse graph}
We observe that an OCR token is informative for answer prediction if it is close to one or more objects, and vice versa for {an object}. This observation prompts us to judge the necessity of the relations between objects and tokens through spatial measurements. We try to cut off useless edges in the graph to facilitate graph relation reasoning, that is, each node in the graph only receives informative messages from its interactive neighbor nodes within a reasonable spatial scale. The implementation details are introduced as follows.

\textbf{OTSG Definition.} There is a sub-graph $\mathcal{G_{VT}}$ = $\left\{\mathcal N,\mathcal{E_{VT}}\right\}$ $\subset$ $\mathcal{G}$, where $\mathcal N$ = $\left({\mathcal{V},\mathcal{T}}\right)$, $\mathcal{E_{VT}}$ = $({\mathcal{E_{V \to T}}}$, ${\mathcal{E_{T \to V}}})$, {where ${\mathcal{E_{V \to T}}}\in$ $\mathbb{R}^{N\times M \times d_e}$, ${\mathcal{E_{T \to V}}}\in$ $\mathbb{R}^{M\times N \times d_e}$.} Taking an edge $\mathcal{E}_{t \to v}$ from OCR token $\tilde t$ to object $\tilde v$ as an example, {the edge feature $\mathcal{E}_{t \to v}$ is performed by} Eq.~\ref{eq:eq5}, and the feature of reverse edge $\mathcal{E}_{v \to t}$ is calculated by Eq.~\ref{eq:eq5} too. 

\textbf{Spatial Relation Pruning.} 
In this case, we deem that the distant object and OCR token have less or even no correlated information. We design the sparsity function with two spatial pruning constraints: \textcircled{1} \textit{Spatial distance}.  It constrains the establishment of relations between objects and OCR tokens when the relative distance $\varDelta d_{tv}$ between token $t_j$ and object $v_i$ is greater than $\theta$*$d_{Img}$, where $d_{Img}$ is the diagonal length of image and $\theta$ is a hyperparameter; \textcircled{2} \emph{DIoU} (\emph{Distance-IoU}) \cite{zheng2020distance} {is a good way to measure the spatial distance and geometric similarity of two boundary boxes. It is formulated in Eq.~\ref{eq:eq10}. Here, we use \emph{DIoU} as a pruning criterion and its value is required to be greater than $\theta$.} In OTSG, the greater the overlap area of the object-OCR token pair is, the closer their relation is.
\begin{equation}{
DIoU = IoU - \frac{{{\rho }^2\left( {n_i,n_j} \right)}}{{{c^2}}},
}
\label{eq:eq10}
\end{equation}
where $\rho(\cdot)$ denotes the Euclidean distance and $c$ is the smallest diagonal length of the bounding box covering both the two nodes $n_i$ and $n_j$. ${IoU}$ is the spatial criterion of \emph{Intersection over Union} \cite{Ren2015FasterRT}. 

We use the \emph{sparsity function} to update the edge set $\mathcal{E_{T \to V}}$ of $\mathcal{G_{VT}}$ as follows.
\begin{equation}{
\mathcal{{E}}_{t_j \to v_i}  = \left\{
\begin{aligned}
&  \mathcal{E}_{t_j \to v_i}, if \ \varDelta d_{tv} \leqslant \theta*d_{Img}\ {\rm{or}} \ DIoU \geqslant \theta;\\
& [0, 0, 0, 0, 0], \ else, \\
\end{aligned}
\right.}
\label{eq:eq7}
\end{equation}
where $\varDelta d_{tv}={\rho}({t_j, v_i})$. Following the spatial setting in previous work \cite{yao2018exploring, li2019relation, yang2020relationship}, we keep the default setting of $\theta =$ 0.5.

\textbf{Graph Inference.} After relation pruning, we implement the graph learning to update all the object and token node features of sub-graph $\mathcal{G_{VT}}$. 
Taking an object node in $\mathcal V$ as an example, we perform a question-guided correlation between object and OCR token. Specifically,  {we calculate the relation matrix $\mathcal{A_{T \to V}} \in\mathbb{R}^{N\times M}$ for message passing from OCR tokens to objects in $\mathcal{G_{VT}}$.
We encapsulate this process as a unified $MP(\cdot)$ function below.
\begin{equation}
\begin{aligned}
&\mathcal{A_{T \to V}} = MP(\mathcal{{G}_{VT}}, \mathcal{{E}_{T \rightarrow V}}, \mathcal{Q}) \\
\Leftrightarrow
&\left\{
\begin{aligned}
& q = \sum\limits_{i = 1}^K{softmax(\bm{W_{q_i}}q_i) \cdot {q}_i}; \\
& a = tanh\left( {\bm{W_{e}}}{\mathcal{{E}_{T \rightarrow V}}+ \bm{W_{q}}{q}} \right); \\
& \mathcal{A_{T \to V}} = softmax\left(\bm{W_a}a\right),
\end{aligned}\right.
\end{aligned}
\label{eq:eq8}
\end{equation}
}
where $\bm{W_{q_i}}$, $\bm{W_{e}}$, $\bm{W_{q}}$, $\bm{W_{a}}$ are learnable parameters. $\mathcal{A_{T \to V}}$ is a similarity matrix, where its element ${\mathcal{A_{T \to V}}}_{ij}$ denotes the correlation weight of two nodes $v_i$ and $t_j$ , whose value is in the range of (0,1).

Then, {we update the object node set $\mathcal{V}$ with the edge set $\mathcal{E_{T \to V}}$ and the relation matrix $\mathcal{A_{T \to V}}$.} Concretely, the node $v_i$ is updated by receiving the messages from its ``connected" token neighbors in the sparse sub-graph. Please note that the token neighbors only exists in the edges $\mathcal{{E}_{T \rightarrow V}}$ updated by Eq. \ref{eq:eq7} rather than all the tokens. {We encapsulate the graph inference process as a unified $GIN(\cdot)$ function below.
\begin{equation}
\begin{aligned}
& \mathcal{V}' = GIN(\mathcal{{G}_{VT}}, \mathcal{E_{T \to V}}, \mathcal{V}, \mathcal{T}, \mathcal{A_{T \to V}}) \\
\Leftrightarrow
& \left\{
\begin{aligned}
& \mathcal{E}'_{\mathcal{T \to V}}=\mathcal{A_{T \to V}} \cdot \bm{W_{\mathcal{E}}} \mathcal{{E}_{\mathcal{T \to V}}}; \\
& \mathcal{M}_{\mathcal{T \to V}} = \bm{W_{\mathcal{T}}} \mathcal{A_{T \to V}}\mathcal{T}; \\
& \mathcal{V}' = {\bm{W_{\mathcal V}}}\mathcal{V} + \bm{W_{\mathcal{E}'}}\mathcal{E'_{T \to V}}+  {\bm{W_{\mathcal M}}{\mathcal{M}_{\mathcal{T \to V}}}},
\end{aligned}\right.
\end{aligned}
\label{eq:eq9}
\end{equation}
}
where $\bm{W_{\mathcal{E}}}$, $\bm{W_{\mathcal{T}}}$, $\bm{W_{\mathcal V}}$, $\bm{W_{\mathcal{E}'}}$, $\bm{W_{\mathcal M}}$ are learnable parameters. We obtain a new object representation ${\mathcal{V}'}$$=$$\{v'_i\}_{i=1}^N  \in \mathbb{R}^{N \times d}$. 

In the same way, we obtain the reverse relation $\mathcal{{E}_{V \rightarrow T}}$ and update any OCR token node in $\mathcal{T}$ by receiving messages from its connected object nodes. As a result, we obtain a new OCR token representation ${\mathcal{T}'}$$=$$\{t'_j\}_{j=1}^M\in \mathbb{R}^{M \times d}$.

\subsubsection{\textbf{Object-Based Sparse Graph (OSG)}}
\label{subsec:Object-based Sparse Graph}
After the above double-sided verification of object and OCR token occurrences, here we narrow the range of relation learning in each visual entity space. In this part, we study the object-object relations in the graph. The objects are always densely detected by the pre-trained detection models (such as Faster RCNN \cite{Ren2015FasterRT} in this work). Superfluous bounding boxes with different sizes covering a same entity often appear, but they are taken as different objects. By observation, these close-by objects have similar geometric sizes and visual appearances, their high similarities result in the strongly intensified relations between them, which may cause imbalanced relations and negatively affect the answer reasoning. To achieve effective visual reasoning, we try to reduce this interference. 

\textbf{OSG Definition.} In this part, we focus on the object entities. Here is a sub-graph {$\mathcal{G_{V}}$ = $\left\{\mathcal{N_{V}},\mathcal{E_{V}}\right\}$}, 
$\mathcal{N_{V}}$ = $\mathcal{{V}'}$, $\mathcal{E_{V}}$ = ${\mathcal{E_{V \to V}}}$ $\in$ $\mathbb{R}^{N \times N \times d_e}$, 
where the edges $\mathcal{E}_{v_i \to v_j}$ and $\mathcal{E}_{v_j \to v_i}$ are also calculated by Eq.~\ref{eq:eq5}. 

\textbf{Spatial Relation Pruning.} 
To address the redundant relations of these similar and close-by objects, {we employ \emph{DIoU} \cite{zheng2020distance} and \emph{spatial distance} again as spatial pruning criteria. }
If DIoU is greater than $\epsilon$, it means that two close objects exist and the connected edge between them has to be cut off. We use the \emph{sparsity function} to update the edge set $\mathcal{E_{V \to V}}$ of $\mathcal{G_{V}}$ below.
\begin{equation}{
\mathcal{{E}}_{v_j \to v_i}  = \left\{
\begin{aligned}
& \mathcal{E}_{v_j \to v_i}, if \  \varDelta d_{v} \!\leqslant\! \theta*d_{Img}\ {\rm{and}} \text{ $DIoU \!\leqslant\! \epsilon$};\\
& [0, 0, 0, 0, 0], else, \\
\end{aligned}                  
\right.}
\label{eq:eq11}
\end{equation}
where $\epsilon$ is a hyperparameter. We set $\varDelta d_v =\rho( {v'_i, v'_j} )$ and $\theta = $0.5 as the same in Eq.~\ref{eq:eq7}.

\textbf{Graph Inference.}
As the same to the above sub-graph $\mathcal {G_{VT}}$, we perform the question-guided object graph learning on $\mathcal {G_{V}}$. The relation matrix $\mathcal{A_{V}}\in \mathbb{R}^{N\times N}$ of $\mathcal {G_{V}}$ is calculated below.
{
\begin{equation}{	
\mathcal{A_{V}} = MP(\mathcal{G_{V}}, \mathcal{E_{V}}, \mathcal{Q}) 
}
\label{eq:eq12}
\end{equation}
}

For the object-object relations in $\mathcal{{{E}}_{V \rightarrow V}}$, we update object features from $\mathcal{{V}'}$ to ${\mathcal{V}''}$$=$$\{v''_i\}_{i=1}^N \in \mathbb{R}^{N\times d}$ by message passing from its ``connected" neighbor objects in the sparse sub-graph OSG  (please note that does not refer to all the objects in $\mathcal{V}'$) as follows.
{
\begin{equation}{
 {\mathcal V''} = GIN(\mathcal{G_{V}}, \mathcal{E_{V}}, {\mathcal V'}, {\mathcal V'}, \mathcal{A_{V}})
}
\label{eq:eq13}
\end{equation}
}

\subsubsection{\textbf{Token-Based Sparse Graph (TSG)}}
\label{subsec:Token-based Sparse Graph}
Here, we discuss the OCR-OCR tokens relations. In the previous work \cite{han2020finding,jin2021ruart,wang2020general,wang2020multimodal,wang2021improving}, scene text recognition has been proven to be significant for TextVQA. Unlike object detection, OCR tokens are detected in relatively small numbers and are independently dispersed, with few or no overlapping regions in the image. As shown in Fig.~\ref{Figure 3}, a small overlap area exists between $ t_i$ ``Jean-Paul" and $t_j$ ``Sartre". $t_i$ and $t_j$ are semantically relevant as they compose {the} name of {a} famous philosopher, where the distant $t_k$ ``Boston" is a city name. The close-by tokens are much more semantic-relevant than distant tokens. Under this consideration, we design a pruning rule for OCR tokens as below.

\textbf{TSG Definition.} The OCR tokens sub-graph is denoted as ${\mathcal{G_{T}} = \left\{\mathcal{N_T},\mathcal{E_{T}}\right\}}$, $\mathcal{N_T}$ = $\mathcal{{T}'}$, $\mathcal{E_{T}}$ = ${\mathcal{E_{T \to T}}}$ $\in$ $\mathbb{R}^{M \times M \times d_e}$, where the edges $\mathcal{E}_{t_i \to t_j}$ and $\mathcal{E}_{t_j \to t_i}$ are also calculated by Eq.~\ref{eq:eq5}. 

\textbf{Spatial Relation Pruning.}
Taking $\mathcal{E}_{t_j \to t_i}$ as an example, we constrain it with the following factors: 
\textcircled{1} \emph{Spatial distance}. It requires that $\varDelta d_t\!\leqslant \alpha\!*\!d_{t_i}$, where $\varDelta d_t$ denotes the shortest bounding box distance between tokens $t'_i$ and $t'_j$, $d_{t_i}$ represents the diagonal length of $t'_i$'s bounding box, and $\alpha$ is a hyperparameter. Opposite to the redundancy of close-by objects, close-by tokens have a close relation. Thus, we cut off the $\mathcal{E}_{t_j \to t_i}$ if it does not meet this condition. \textcircled{2} \emph{Geometric size}. We restrict the heights of close-by OCR tokens. By observation, the tokens gathering together to describe a phrase or sentence are often in similar font sizes. Another fact is that different font sizes reflect different semantic importance. We attempt to find out the scene texts at the same semantic level in the image. We set $\beta*h_i\leqslant h_j\leqslant \gamma*h_i$, where $h_i$ and $h_j$ denote the normalized height of $t'_i$ and $t'_j$ in the  image, and $\mathnormal{\beta}, \mathnormal{\gamma}$ are hyperparameters. \textcircled{3} \emph{Overlap area.} To make sure the readability of the OCR tokens, various OCR systems make efforts to output less or no overlap tokens. We follow this rule to calculate an overlap ratio $\varDelta_A$ = $max(\frac{A_{ij}}{A_i},\frac{A_{ij}}{A_j})$ of tokens $t'_i$ and $t'_j$ and set it less than threshold $\delta$, where $A_i$ and $A_j$ are bounding box areas of ${t'_i}$ and ${t'_j}$, respectively. $A_{ij}$ is the intersection area between the bounding boxes of ${t'_i}$ and ${t'_j}$. To summarize, we use the \emph{sparsity function} to update the edge set $\mathcal{E_{T \to T}}$ of $\mathcal{G_{T}}$ as follows.
\begin{equation}{
\mathcal{{E}}_{t_j \to t_i} \! = \! \left\{
\begin{aligned}
& \mathcal{E}_{t_j \to t_i}, if \varDelta d_t \!\leqslant\! \alpha \!*\! d_{t_i}, h_j \!\!\in \!\! [\beta \!, \gamma ]\!*\! h_i, {\rm{and}} \ \varDelta_A \! \leqslant \! \delta; \\
& [0, 0, 0, 0, 0], \ else, \\
\end{aligned}
\right.}
\label{eq:eq14}
\end{equation}
where $\alpha$, $\beta$ and $\gamma$ are hyperparameters. 
We set the threshold $\delta$ = 0.5 following the OCR spatial setting~\cite{wang2020multimodal}.

\textbf{Graph Inference.} 
Up to now, the sparse sub-graph $\mathcal {G_{T}}$ referring to OCR token nodes is built. The token graph learning process of $\mathcal{G_{T}}$ is performed the same as $\mathcal {G_{T}}$. We first calculate the relation matrix $\mathcal{A_{T}} = MP(\mathcal{G_{T}}, \mathcal{E_{T}}, \mathcal{Q}) \in \mathbb{R}^{M\times M}$ to perform the question-guided message passing among tokens. Any OCR token node $t'_j$ is updated by its ``connected" neighbor tokens. 
At last, we update the token node set by the unified graph inference function $\mathcal{T}'' = GIN({\mathcal{G_T}}, \mathcal{E_{T}}, \mathcal{T'}, \mathcal{T'}, \mathcal{A_{T}})\in\mathbb{R}^{M\times d}$. 

\subsection{Answer Generation}
\label{subsec:Answer Generation Module}
\subsubsection{\textbf{Answer Prediction}} Based on the final output node representations ${\mathcal{V}''}$ and ${\mathcal{T}}''$, we adopt an available text generator \cite{hu2020iterative,liu2020cascade} for answer prediction, which is made up of a transformer and two classifiers---the object classifier $\psi_o$ and the OCR token classifier  $\psi_t$. We concatenate ${\mathcal Q}$, ${\mathcal{V}''}$, ${\mathcal{T}''}$  and a hidden state $o \in \mathbb{R}^d$ and input them into a transformer module as follows.
\begin{equation}
\begin{aligned}
& [{\mathcal{\check Q}}, {\mathcal{\check V}}, {\mathcal{\check T}}, {\check o}] = \varPsi([{\bm{W_{\mathcal{Q}}}\mathcal Q}, {\bm{W_{\mathcal{V}''}}\mathcal{V}''}, {\bm{W_{\mathcal{T}''}}\mathcal{T}''}, \bm{W_o}o]), \\
\end{aligned}
\label{eq:eq15}
\end{equation}
where $\bm{W_{\mathcal{Q}}}, \bm{W_{\mathcal{V}''}}, \bm{W_{\mathcal{T}''}}, \bm{W_o}$ are learnable parameters, $\varPsi(\cdot)$ is a four-layer transformer, and $o$ is initialized by positional embedding \cite{hu2020iterative}.
We perform Eq.~\ref{eq:eq15} $L$ times, thus we obtain a generated sequence $\mathcal{\check O} = [\check o_1$, $ \cdots$, $ \check o_L]\in\mathbb{R}^{d\times L}$.

This part can be regarded as a multi-label classification. At each $l$-th decoding time, the object classifier $\psi_o$ is realized by a simple linear layer and predicts the probability score $y^o_l$ over a pre-set object vocabulary. Another classifier, $\psi_t$, is proposed to compute the token score $y^t_l$ by the dot product of the generated $\check o_l$ and OCR tokens $\mathcal{ \check T}$, where $\mathcal{ \check T}$ is a dynamic OCR token set detected in each image. Formally, the predicted scores $y^o_l$ and $y^t_l$ are calculated as below:
\begin{equation}{	
\left\{\begin{matrix}
\begin{aligned}
y^o_l &= \bm{W_l^{o}} \check o_l + \bm{b_l^{o}}; \\
y^t_l &= (\bm{W_l^{t1}}\mathcal{ \check T} + \bm{b_l^{t1}})^\top(\bm{W_l^{t2}}\check o_l + \bm{b_l^{t2}}), \\
\end{aligned}
\end{matrix}\right.
}
\label{eq:eq16}
\end{equation}
where $\bm{W_l^{o}}$, $\bm{W_l^{t1}}$, $\bm{W_l^{t2}}$ are learnable parameters and $\bm{ b_l^{o}}$, $\bm{b_l^{t1}}$, $\bm{b_l^{t2}}$ are scalar parameters at the $l$-th timestamp.

At last, we implement the $argmax$ function on $y^o_l$ and $y^t_l$ to predict the answer word $y_l^{pred}$. Thus, the answer sentence with length $L$, {$y^{pred} = \{y_l^{pred}\}_{l=1}^L$} is represented as:
\begin{equation}
y_l^{pred}=argmax([y^o_l,y^t_l]),
\label{eq:eq17}
\end{equation}

\subsubsection{\textbf{Training Loss}}
Following the previous work \cite{hu2020iterative,liu2020cascade}, binary cross-entropy loss $\mathcal L_{bce}$ is widely used for TextVQA. In real applications, the utterly correct answer sentence is expected but rarely occurs, whereas the answer with semantically similar words is acceptable. 
{Following \cite{liu2020cascade}}, a new auxiliary policy gradient loss $\mathcal L_{pg}$ based on ANLS (\emph{Average Normalized Levenshtein Similarity} (stated in Eq.~\ref{eq:eq20} \cite{biten2019scene}) is introduced into this task \cite{singh2019towards}. The ANLS measures the character-level composition similarity between the predicted and ground-truth answers  as follows.

\begin{equation}{	
\left\{\begin{matrix}
\begin{aligned}
\mathcal L_{bce} &= - {y^{gt}} \log (\sigma ({y^{pred}})) - (1-y^{gt}) \log (1 - \sigma (y^{pred}));\\
\mathcal L_{pg} &= - \log (\sigma({y^{pred}}))\cdot
{\rm{ANLS}}(y^{gt}, y^{pred}),
\end{aligned}
\end{matrix}\right.
}
\label{eq:eq18}
\end{equation}
where $\sigma$($\cdot$) is sigmiod function, {$y^{gt}$} is the ground-truth. 

By combing $\mathcal L_{bce}$ and $\mathcal L_{pg}$, the total loss is formulated as follows.
\begin{equation}
\mathcal L = \mathcal L_{bce} + \lambda  \mathcal L_{pg},
\label{eq:eq19}
\end{equation}
where $\lambda$ is a trade-off hyperparameter.

\section{Experiment}
\label{subsec:Experiment}
\subsection{Datasets}
Experiments are conducted on two benchmark datasets of text-based visual question answering. 
\subsubsection{\textbf{TextVQA} \cite{singh2019towards}} In this dataset, images are collected from Open Images v3 \cite{krasin2017openimages}. It contains 28,408 images and 45,336 questions, which consists of a training set of 21,953 images and 34,602 question-answer (QA) pairs, a validation set of 3,166 images and 5,000 QA pairs, and a test set of 3,289 images and 5,734 QA pairs \cite{hu2020iterative}. For each image, there are about one or two QA pairs. The average lengths of question and answer are 7.18 and 1.70 words, respectively. 
{Questions in this dataset are interested in} visual objects or scene texts in the images. Up to 39\% (about 18K) of answers do not contain any OCR token.

\subsubsection{\textbf{ST-VQA} \cite{biten2019scene}}
The dataset comprises 23,038 images and 31,791 questions, which is collected from six different datasets of ICDAR 2013 \cite{karatzas2013icdar}, ICDAR 2015 \cite{karatzas2015icdar}, ImageNet \cite{deng2009imagenet}, VizWiz \cite{gurari2018vizwiz}, IIIT Scene Text Retrieval \cite{mishra2013image}, Visual Genome \cite{krishna2017visual}, and COCO-Text \cite{veit2016coco}. Following the protocol \cite{hu2020iterative}, this dataset is divided into the {train/val/test sets of 17,028/1,893/2,971 images and 23,446/2,628/4,070} QA pairs, respectively. Compared with the TextVQA dataset, ST-VQA more emphasizes scene texts as all the questions have to be answered with scene texts. Each image contains more than two scene texts, regardless of whether or not they can be detected by the OCR systems. {ST-VQA introduces three novel tasks, namely the strongly contextualized task (Task 1), the weakly contextualized task (Task 2), and the open vocabulary task (Task 3). Specifically, Task 1 provides a dynamic candidate dictionary of 100 words defined for per image; Task 2 provides a fixed answer dictionary of 30, 000 words for the whole dataset; following \cite{hu2020iterative}, Task 3 provides a fixed answer dictionary of 5,000 words for the whole dataset.}


\subsection{Evaluation Metrics}
\label{subsection:Evaluation Metrics}
We adopt accuracy (\textbf{\emph{Acc}}) as a basic evaluation metric following \cite{hu2020iterative,kant2020spatially,Zhu2021SimpleIN}. Each question in the TextVQA dataset has ten human-annotated answers, and the final accuracy is the average score over these ten answers. As for ST-VQA \cite{biten2019scene}, there is a new evaluation metric \textbf{\emph{ANLS}} which measures the frequency of correct words in each generated answer {as defined in Eq. \ref{eq:eq20}}. \textbf{\emph{ANLS}} is calculated in term of the correct words, while \textbf{\emph{Acc}} is calculated in term of the complete sentence. 
\begin{equation}{
{ANLS(y^{pred},y^{gt}) = 1 -\frac{NL(y^{pred},y^{gt})}{max(|y^{pred}|,|y^{gt}|)}.}
}
\label{eq:eq20}
\end{equation}
where \emph{NL} denotes the normalized Levenshtein distance \cite{levenshtein1966binary} , {$y^{pred}$ and $y^{gt}$} denote the predicted and ground-truth answers, respectively. As set in ST-VQA \cite{biten2019scene}, the score \emph{{ANLS}} is set to 0 if it is below 0.5. 

\begin{table}[t]
\centering
\fontsize{6}{8}\selectfont
\caption{The statistics of OCR token output by different OCR systems on TextVQA and ST-VQA datasets. \emph{Total} represents the sum number of OCR tokens. \emph{Mean}, \emph{Min} and \emph{Max} represent the average, minimum, and maximum numbers of OCR tokens per image. \emph{Min} = 0 occurs in the cases of the blur, partially occluded scene texts, or scene text in illegible fancy fonts, \emph{etc.} \emph{Max} = 2694 occurs in case of scene text for reading, such as book pages. In our experiments, we choose $M$ = 50 OCR tokens for experiments. }
\label{tab:table9}
\resizebox{0.48\textwidth}{!}{
\begin{tabular}{llcccc}
\toprule
\textbf{\makecell{Dataset}}&\textbf{\makecell[c]{OCR System}}&\textbf{\makecell{Total}} &\textbf{\makecell{Mean}} & \textbf{\makecell{Min}} & \textbf{\makecell{Max}}  \\
\hline
\multirow{4}{*}{{\makecell[c]{TextVQA}}}  
& Rosetta-en  & 566,824 & 12.50 & 0  &  100   \\
& SBD-Trans   & 914,521 & 20.17 & 0 &  247 \\
& Google-OCR  & 1,304,155 & 28.77 & 0  &  2,657  \\
& Microsoft-OCR  & 1,419,941 & 31.32 & 0 & 2,694  \\
\hline
\multirow{4}{*}{{\makecell[c]{ST-VQA}}}  
& Rosetta-en  & 226,643 & 7.52 & 0  & 99   \\
& SBD-Trans   & 266,947 & 8.86 & 0  &  100  \\
& Google-OCR  & 292,196  & 9.69 & 0  &  477  \\
& Microsoft-OCR  & 359,137 & 11.91 & 0  & 358 \\
\bottomrule
\end{tabular}}
\end{table}

\subsection{Implementation Details}
For feature extraction, we follow the experimental settings of
\cite{singh2019towards} and \cite{biten2019scene}. We use the Faster R-CNN \cite{Ren2015FasterRT} pre-trained on Visual Genome dataset \cite{krishna2017visual} to detect objects. Each object has a 2048-$dim$ appearance feature and a 4-$dim$ boundary box feature. They are encoded by a separate fully connected layer and then added into a 768-$dim$ vector and used as an original object feature. As for OCR tokens, we {conduct experiments on four OCR systems, \ie,} Rosetta-en \cite{borisyuk2018rosetta}, SBD-Trans \cite{Liu2019OmnidirectionalST}, Google-OCR \textsuperscript{\ref {fontnote1}}, and Microsoft-OCR \textsuperscript{\ref {fontnote2}}. Each OCR token feature consists of four aspects, including 300-$dim$ FastText feature \cite{bojanowski2017enriching}, 604-$dim$ PHOC (pyramidal histogram of characters) feature \cite{almazan2014word}, 2048-$dim$ appearance feature{,} and 4-$dim$ bounding box feature. Following the previous work for TextVQA \cite{hu2020iterative, liu2020cascade, zhang2021position}, {all features are encoded by a separate fully connected layer and then added to obtain a 768-$dim$ vector.} Table~\ref{tab:table9} shows the statistics of OCR tokens detected on the TextVQA and ST-VQA datasets. 
In our experiments, we choose $N$ = 100 objects with the best probabilities and $M$ = 50 OCR tokens following \cite{Zhu2021SimpleIN, hu2020iterative, zeng2021beyond, zhang2021position}.

About the other experiment setups, each question sentence is truncated with the length $K$ = 20 and equipped with 768-$dim$ word embedding. In this work, we perform a two-layer transformer with 12 heads for feature preparation in Sec.~\ref{subsec:Feature Preparation} and a four-layer transformer with 12 heads for answer decoding in Sec.~\ref{subsec:Answer Generation Module}. The maximum length of the output answer $L$ = 12, and the trade-off parameter in the total loss objective is set to $\lambda$ = 1. The threshold $\theta$ is set to 0.5 following the setting of spatial exploration \cite{yao2018exploring}. We set empirical parameters with $\epsilon$ = 0.3, $\alpha$ = 5, $\beta$ = 0.3 and $\gamma$ = 2.0. We choose Adam as the optimizer, and the learning rate is set to 1e$-$4. During training, we multiply the learning rate by 0.1 at 10,000 and 21,000 iterations, respectively, for a total of 24,000 iterations.

\begin{table}
\centering
\caption{Main comparison on TextVQA dataset. The grey block marks the pre-training techniques. {*} denotes that the model is jointly trained with extra pre-training tasks of MLM, ITM, and RPP. {\dag} denotes jointly pre-training with the question-visual grounding task. VG denotes the Visual Genome dataset \cite{krishna2017visual}. {{``F''} denotes full relations, and {``S''} denotes sparse relations.} }
\vspace{-0.3cm}
\label{tab:table1}
\fontsize{7}{9}\selectfont
\resizebox{0.48\textwidth}{!}{
\begin{tabular}{c|c|c|cc}
\Xhline{1.2pt}
\textbf{Method} & \makecell{\textbf{OCR  System}} &  \makecell{\textbf{Extra Data}} & \textbf{\makecell{Val  \\ Acc}} & \textbf{\makecell{Test \\ Acc}} \\
\Xhline{1.2pt}
\multicolumn{5}{c}{\textbf{Attention-based models}} \\
\hline
LoRRA \cite{singh2019towards} {(F)} & Rosetta-ml & - & 26.56 & 27.63  \\
SSBaseline \cite{Zhu2021SimpleIN} {(F)} & Rosetta-en &  - & 40.38 & 40.92   \\
SSBaseline \cite{Zhu2021SimpleIN} {(F)} & SBD-Trans &  - & 43.95 & 44.72   \\
SSBaseline \cite{Zhu2021SimpleIN} {(F)} & SBD-Trans & ST-VQA & 45.53 & 45.66   \\
\hline
\multicolumn{5}{c}{\textbf{Transformer-based models}} \\
\hline
M4C \cite{hu2020iterative} {(F)} & Rosetta-en & -  & 39.40 &  39.01  \\
M4C \cite{hu2020iterative} {(F)} & Rosetta-en & ST-VQA & 40.55 &  40.46  \\
LaAP-Net \cite{han2020finding} {(F)} & Rosetta-en &  - & 40.68 &  40.54  \\
LaAP-Net \cite{han2020finding} {(F)} & Rosetta-en & ST-VQA & 41.02  & 40.54   \\
PAT \cite{zhang2021position} {(F)} &  Google-OCR &  - & 42.80 & 43.41   \\
\cellcolor[gray]{0.9}TAP$^{*}$ \cite{yang2021tap} {(F)} &  \cellcolor[gray]{0.9}Microsoft-OCR &  \cellcolor[gray]{0.9}- & \cellcolor[gray]{0.9}49.91 & \cellcolor[gray]{0.9}49.71  \\
\cellcolor[gray]{0.9}TAP$^{*}$ \cite{yang2021tap} {(F)} &  \cellcolor[gray]{0.9}Microsoft-OCR & \cellcolor[gray]{0.9}ST-VQA & \cellcolor[gray]{0.9}50.57 & \cellcolor[gray]{0.9}50.71  \\
\cellcolor[gray]{0.9}LOGOS$^{\dag}$ \cite{Lu2021LocalizeGA} {(F)} &  \cellcolor[gray]{0.9}Microsoft-OCR & \cellcolor[gray]{0.9}VG & \cellcolor[gray]{0.9}50.79 & \cellcolor[gray]{0.9}50.65   \\
\cellcolor[gray]{0.9}LOGOS$^{\dag}$ \cite{Lu2021LocalizeGA} {(F)} &  \cellcolor[gray]{0.9}Microsoft-OCR &  \cellcolor[gray]{0.9}ST-VQA, VG & \cellcolor[gray]{0.9}51.53 &  \cellcolor[gray]{0.9}51.08  \\
\hline
\multicolumn{5}{c}{\textbf{Representation learning models}} \\
\hline
BOV \cite{zeng2021beyond} {(F)} & Rosetta-en &  - &  {40.90} &  {41.23}   \\
BOV \cite{zeng2021beyond} {(F)} & SBD-Trans &  - &  {44.87} &  {45.63}   \\
BOV \cite{zeng2021beyond} {(F)} & SBD-Trans & ST-VQA &  {46.24} & \underline{46.96}   \\
\hline
\multicolumn{5}{c}{\textbf{Graph-based models}} \\
\hline 
MM-GNN \cite{gao2020multi} {(F)} & Rosetta-ml & -  & 31.44 &  31.10 \\
CRN \cite{liu2020cascade} {(F)} &  Rosetta-en & -  & 40.39 &  40.96  \\
SA-M4C \cite{kant2020spatially} {(F)} &  Google-OCR & -  & 43.90 & -   \\
SA-M4C \cite{kant2020spatially} {(F)} &  Google-OCR & ST-VQA & 45.40 &  44.60  \\
SMA \cite{gao2020structured} {(S)} & Rosetta-en & -  & 40.05 &  40.66  \\
SMA \cite{gao2020structured} {(S)} &  SBD-Trans & -  & 44.58 &  45.51  \\
SMA \cite{gao2020structured} {(S)} & SBD-Trans & ST-VQA & 44.58 &  45.51  \\
{SSGN (Ours)} {(S)} & {Rosetta-en} & -  & {42.00}  & {41.60} \\
{SSGN (Ours)} {(S)} & {SBD-Trans} & - & {45.72} &  {46.63}  \\
{SSGN (Ours)} {(S)} & {SBD-Trans} & {ST-VQA}  & \textbf{46.96} &  {46.63}   \\
{SSGN (Ours)} {(S)} & {Google-OCR} & {ST-VQA} & {45.45} & {45.42}  \\
\textbf{SSGN (Ours)} {(S)} & \textbf{Microsoft-OCR} & \textbf{ST-VQA}  & \underline{46.85} &  \textbf{47.16}   \\
\Xhline{1.2pt}
\end{tabular}}
\vspace{-0.3cm}
\end{table}

\subsection{Comparison with State-of-the-art Results}
In this subsection, we compare the proposed method with the state-of-the-art approaches---\textbf{Attention-based models} (LoRRA \cite{singh2019towards}, SSBaseline \cite{Zhu2021SimpleIN}), \textbf{Transformer-based models} (M4C \cite{hu2020iterative}, LaAP-Net \cite{han2020finding}, PAT \cite{zhang2021position}, TAP \cite{yang2021tap}, LOGOS \cite{Lu2021LocalizeGA}), \textbf{Representation learning models} (BOV \cite{zeng2021beyond}), and \textbf{Graph-based models} (MM-GNN \cite{gao2020multi}, CRN \cite{liu2020cascade}, SA-M4C \cite{kant2020spatially}, SMA \cite{gao2020structured}).

\subsubsection{\textbf{Results on TextVQA}}
As the experimental results shown in Table~\ref{tab:table1}, the proposed \textbf{SSGN} achieves a promising performance {compared to state-of-the-art methods.} Compared with \textbf{SSBaseline} \cite{singh2019towards} (an attention-based model), under the same setup of SBD-Trans features and ST-VQA \cite{biten2019scene} training data, \textbf{SSGN} improves 1.43\% on the val set and 0.97\% on the test set. Compared with \textbf{BOV} \cite{zeng2021beyond} {(a representation learning model)} based on SBD-Trans features, \textbf{SSGN (Ours)} improves 0.85\% on the val set and 1\% on the test set, and when taking ST-VQA as extra training data, \textbf{SSGN (Ours)} further improves 1.24\% on the val set. Compared with \textbf{LaAP-Net} \cite{han2020finding} (a transformer-based model without pre-trained techniques), \textbf{SSGN (Ours)} improves 1.32\% on the val set and 1.06\% on the test set. By introducing pre-trained techniques such as MLM (masked language modeling), ITM (image-text matching), RPP (relative position prediction) or large-scale extra data such as the Visual Genome dataset (which includes 108,000 images, 5.4 million descriptions, 1.7 million QA pairs and 2.3 million relation annotations) \cite{krishna2017visual}, \textbf{TAP} \cite{yang2021tap} and \textbf{LOGOS} \cite{Lu2021LocalizeGA} obtains higher performance than ours.

\begin{table}
\caption{Main comparison on three tasks of ST-VQA dataset. {{``F''} denotes full relations, and {``S''} denotes sparse relations.}} 
\vspace{-0.3cm}
\label{tab:table2}
\scriptsize
\centering
\setlength{\tabcolsep}{.2em}
\renewcommand{\arraystretch}{1.2}
\resizebox{\columnwidth}{!}{
\begin{tabular}{c|c|c|c|c|ccc}
\Xhline{1.2pt}
\multirow{3}{*}{\textbf{Method}}&\multirow{3}{*}{\textbf{\makecell{OCR System}}}&\multirow{3}{*}{\textbf{\makecell{Extra Data}}}  
&\textbf{\makecell{{Task 1}}} & \textbf{\makecell{{Task 2}}}  & \multicolumn{3}{c}{\textbf{Task 3}} \\ \cline{4-8} &&&\textbf{\makecell[c]{{Test} \\{ANLS}}} & \textbf{\makecell[c]{{Test} \\ {ANLS}}}& \textbf{\makecell[c]{Val \\ Acc}} & \textbf{\makecell[c]{Val \\ ANLS}} & \textbf{\makecell[c]{Test \\ ANLS}}  \\
\Xhline{1.2pt}
\multicolumn{8}{c}{\textbf{Attention-based models}} \\
\hline
SAN+STR \cite{biten2019scene} {(F)} & {-} & - & 0.135 & 0.135 & 10.46 &  - & {0.135}  \\
VTA \cite{Biten2019ICDAR2C} {(F)} & {-} & - & - & 0.279 & 18.13 &  - & {0.282} \\
SSBaseline \cite{Zhu2021SimpleIN} {(F)} & {SBD-Trans} & - & 0.506 & 0.505 & - &  - & {0.509} \\
\hline
\multicolumn{8}{c}{\textbf{Transformer-based models}} \\
\hline
M4C \cite{hu2020iterative} {(F)} & Rosetta-en & - & - & - & 38.05 & 0.472  & 0.462 \\
LaAP-Net \cite{han2020finding} {(F)} & Rosetta-en & - & - & - & 39.74 & 0.497 & 0.485 \\
PAT \cite{zhang2021position} {(F)} & Google-OCR & - & - & - & 41.10 & - & 0.508 \\
\cellcolor[gray]{0.9}TAP$^{*}$ \cite{yang2021tap} {(F)} &  \cellcolor[gray]{0.9}Microsoft-OCR &  \cellcolor[gray]{0.9}- &\cellcolor[gray]{0.9}- & \cellcolor[gray]{0.9}- & \cellcolor[gray]{0.9}45.29 &  \cellcolor[gray]{0.9}0.551 & \cellcolor[gray]{0.9}0.543 \\
\cellcolor[gray]{0.9}LOGOS$^{\dag}$ \cite{Lu2021LocalizeGA} {(F)} &  \cellcolor[gray]{0.9}Microsoft-OCR & \cellcolor[gray]{0.9}VG & \cellcolor[gray]{0.9}- & \cellcolor[gray]{0.9}- & \cellcolor[gray]{0.9}44.10 & \cellcolor[gray]{0.9}0.535 & \cellcolor[gray]{0.9}0.522 \\ 
\cellcolor[gray]{0.9}LOGOS$^{\dag}$ \cite{Lu2021LocalizeGA} {(F)} &  \cellcolor[gray]{0.9}Microsoft-OCR & \cellcolor[gray]{0.9}TextVQA, VG & \cellcolor[gray]{0.9}- & \cellcolor[gray]{0.9}- & \cellcolor[gray]{0.9}\underline{48.63} & \cellcolor[gray]{0.9}\underline{0.581} & \cellcolor[gray]{0.9}\textbf{0.579} \\ 
\hline
\multicolumn{8}{c}{\textbf{Representation learning models}} \\
\hline
BOV \cite{zeng2021beyond} {(F)} & Rosetta-en & - & & & 40.18 &  0.500 & 0.472 \\
\hline
\multicolumn{8}{c}{\textbf{Graph-based models}} \\
\hline
MM-GNN \cite{gao2020multi} {(F)} & Rosetta-ml & - & 0.203 & - & - & - &  0.207 \\
CRN \cite{liu2020cascade} {(F)} & Rosetta-en & -  & - & 0.482 & - & - &  0.483 \\
SA-M4C \cite{kant2020spatially} {(F)} & Google-OCR & - & - & - & {42.23} &  {0.512} & 0.504  \\
SMA \cite{gao2020structured} {(S)} & Rosetta-en & - & - & - & -  & - & 0.486 \\
{SSGN(Ours)} {(S)} & {Rosetta-en} & - & {0.487} & {0.495} & {40.12} & {0.493} & {0.490} \\
{SSGN (Ours)} {(S)} & {SBD-Trans} & - & {0.509} & {0.507} & {42.50} & {0.519} & {0.507} \\
{SSGN (Ours)} {(S)} & {SBD-Trans} & {TextVQA} & \underline{{{0.547}}} &  \underline{{{0.550}}} & 44.72 & 0.548 & 0.535 \\ 
{SSGN (Ours)} {(S)} & {Google-OCR} & {TextVQA} & {0.520} & {0.526} & 44.69 & 0.537 & 0.523 \\
\textbf{SSGN (Ours)} {(S)} & \textbf{Microsoft-OCR} & \textbf{TextVQA} & \textbf{{0.570}} & \textbf{{0.573}} & \textbf{48.81} & \textbf{0.589} & \underline{0.573} \\
\Xhline{1.2pt}
\end{tabular}}
\vspace{-0.3cm}
\end{table}

The fairest comparison exists among the graph-based models. Compared with \textbf{MM-GNN} \cite{gao2020multi} (a total fully-connected graph model), when using Rosetta-en features, our model achieves 10.56\% and 10.50\% improvements on the val and test sets respectively. \textbf{CRN} \cite{liu2020cascade} constructs a fully-connected graph that merely explores the relations between objects and OCR tokens, namely ignoring object-object and token-token relations. Our model improves upon \textbf{CRN} by 1.61\% on the val set and 0.64\% on the test set. \textbf{SA-M4C} \cite{kant2020spatially} introduces a spatial orientation factor into the graph modeling but does not consider the relation redundancy in the graph. {Compared with \textbf{SA-M4C} \cite{kant2020spatially}, our model achieves 0.05\% and 0.82\% improvements on the val and test sets when using Google-OCR features and ST-VQA data for training.} In \textbf{SMA} \cite{gao2020structured}, each node adaptively selects the top-5 nearest neighbors for relation learning. When using the SBD-Trans features and extra training data, our model achieves 2.38\% and 1.12\% improvements on the val set and the test set than \textbf{SMA}.

\subsubsection{\textbf{Results on ST-VQA}}
The questions in the ST-VQA dataset are answered more explicitly by utilizing the OCR tokens in the images than in the TextVQA dataset. {From Table~\ref{tab:table2}, our method shows absolute superiority on the three tasks of ST-VQA dataset. In Task 1, \textbf{SSGN (Ours)} obtains 48.7\% on Test ANLS and surpasses \textbf{MM-GNN} \cite{gao2020multi} by 28.7\%. In Task 2, \textbf{SSGN (Ours)} reaches 49.5\% on Test ANLS and outperforms \textbf{CRN} \cite{liu2020cascade} by 1.3\%. In Task 3, when using Microsoft-OCR, \textbf{SSGN (Ours)} achieves 58.9\% on Val ANLS and 48.81\% on Val Acc, surpassing all other methods on the {val} set. The results show that our model has competitive performance compared with other methods in all three tasks.}

\subsection{Role of Graph Module}
\label{Role of Graph Inference}
In this subsection, we test the role of each graph module in our method. 
As shown in Table~\ref{tab:table3}, \textbf{w/o graph} that removes all the graph modules in our method shows the most severe performance degradation.

\subsubsection{\textbf{Single Sub-Graph}}
We test a single sub-graph and report the experimental results in Table~\ref{tab:table3}. Single \textbf{OTSG} means that we merely save the OTSG graph module and remove the other graph modules in our method. Compared with the full model \textbf{SSGN}, \textbf{OTSG} decreases 1.53\% on the val set and 0.93\% on the test set obviously. This indicates that using object-OCR token interaction alone is not sufficient for answer prediction. Single \textbf{OSG} with object nodes achieves the worst performance, with a drop of 2\% on the val set. The surprising case is the single \textbf{TSG} with OCR token nodes, which decreases slightly by 0.79\% in the val set compared to the full model. This suggests that the OCR tokens indeed play an important role in this task. 

\subsubsection{\textbf{Dual Sub-Graphs}}
We test various combinations of two sub-graphs. Among them, \textbf{OTSG\&TSG} achieves the best performance, with a slight decrease of 0.37\% on the val set compared with the full model \textbf{SSGN}. This result is reasonable because OCR tokens are important in both OTSG and TSG graphs. In contrast, \textbf{OTSG\&OSG} has a substantial decrease of 1.51\% on the val set, and even is much worse than the parallel learning of \textbf{OSG\&TSG} with an accuracy reduction of 0.73\%. We speculate that the object redundancy is much more serious than the OCR tokens. The redundant relations between objects may interfere with the reasoning process.

\begin{table}[t]
\centering
\caption{The performance of graph modules on TextVQA dataset.}
\vspace{-0.3cm}
\label{tab:table3}
\fontsize{10}{12}\selectfont
\resizebox{0.48\textwidth}{!}{
\begin{tabular}{lccccc}
\Xhline{1.5pt}
\textbf{Method} & \textbf{\makecell[c]{OSG}} & \textbf{\makecell[c]{TSG}} & \textbf{\makecell[c]{OTSG}} & \textbf{\makecell[c]{Val Acc}} & \textbf{\makecell[c]{Test Acc}} \\
\Xhline{1.5pt}
w/o graph & - & - & - & 39.92  & 40.51 \\
\hline
OSG & $\checkmark$ & - & - &  40.00 & 41.32 \\
TSG & - & $\checkmark$ & - & \underline{41.21} & \underline{41.51} \\
OTSG & - & - & $\checkmark$ & 40.47  & 40.67 \\
\hline
OSG\&TSG & $\checkmark$ &  $\checkmark$ & - & 41.27 & \underline{41.57}  \\
OTSG\&OSG & $\checkmark$ & - & $\checkmark$ & 40.49 & 41.18 \\
OTSG\&TSG & - & $\checkmark$ & $\checkmark$ & \underline{41.63} & {41.55}  \\			
\hline
OSG\&TSG\&OTSG & {$\checkmark$} & {$\checkmark$} & {$\checkmark$} & 40.14  & 40.42 \\
OSG\&TSG$\rightarrow$OTSG & {$\checkmark$} & {$\checkmark$} & {$\checkmark$} & {41.24} & 41.20 \\
\textbf{OTSG$\rightarrow$OSG\&TSG (Ours)} & \textbf{$\checkmark$} & \textbf{$\checkmark$} & \textbf{$\checkmark$} & \textbf{42.00} & \textbf{41.60} \\
\Xhline{1.5pt}
\end{tabular}}
\vspace{-0.3cm}
\end{table}

\begin{table}
\centering
\fontsize{8}{10}\selectfont
\caption{Ablation studies of graph sparsity on TextVQA dataset.}
\vspace{-0.3cm}
\label{tab:table4}
\resizebox{0.48\textwidth}{!}{
\begin{tabular}{lccccc}
\Xhline{1.2pt}
\textbf{Method} & \textbf{\makecell[c]{OSG}} & \textbf{\makecell[c]{TSG}} & \textbf{\makecell[c]{OTSG}} & \textbf{\makecell[c]{Val Acc}} & \textbf{\makecell[c]{Test Acc}} \\
\Xhline{1.2pt}
w/o sparsity & - &  - & - & 40.99 & 40.42  \\
\hline
w/ OSG sparsity & $\checkmark$ &  - & - & 41.10 & 41.10  \\
w/ TSG sparsity & - & $\checkmark$ & - & 41.08 & 41.08 \\
w/ OTSG sparsity & - & - & $\checkmark$ &  \underline{41.32} & \underline{41.57} \\
\hline
w/o OTSG sparsity & $\checkmark$ & $\checkmark$ & - & 41.43 & 41.32  \\
w/o OSG sparsity & - & $\checkmark$ &$\checkmark$ & 41.34 & 41.30 \\
w/o TSG sparsity & $\checkmark$ & - &  $\checkmark$ & \underline{41.49} & \underline{41.51}  \\
\hline
\textbf{SSGN (Ours)} &  $\checkmark$ & $\checkmark$ &$\checkmark$ & \textbf{42.00} & \textbf{41.60} \\
\Xhline{1.2pt}
\end{tabular}}
\vspace{-0.3cm}
\end{table}

\subsubsection{\textbf{Hierarchical Graph Structure}}
Further, we test the hierarchical graph structure including the following three graph variants: \textcircled{1} A parallel learning of \textbf{OSG\&TSG\&OTSG}. In this case, we concatenate the object features output by OSG and OTSG as ${\mathcal{V}''}$ and the token features output by TSG and OTSG as ${\mathcal{T}''}$ for answer generation, \textcircled{2} \textbf{OTSG$\rightarrow$OSG\&TSG} (Ours), and \textcircled{3} \textbf{{OSG\&TSG$\rightarrow$OTSG}}, a cascade graph learning approach in the opposite order of \textbf{OTSG$\rightarrow$OSG\&TSG}. Among these structures, \textbf{OSG\&TSG\&OTSG} has the worst performance. Compared to \textbf{OTSG$\rightarrow$OSG\&TSG} (Ours), it drops 1.86\% on the val set and 1.18\% on the test set. The \textbf{OSG\&TSG\&OTSG} performs even worse on any combination of dual sub-graphs. This may be because redundant relations are amplified in this three-sub-graph parallel learning. By comparing \textbf{OSG\&TSG$\rightarrow$OTSG} and \textbf{OTSG$\rightarrow$OSG\&TSG}, the latter is clearly more effective. We insist on the validity of \textbf{OTSG$\rightarrow$OSG\&TSG}, which first implements the correlation of object-OCR token and then examines the correlated clues in each visual space individually.

\subsection{Ablation Studies} 
\begin{figure}[t]
\centering
\includegraphics[height=6.8cm,width=8.8cm]{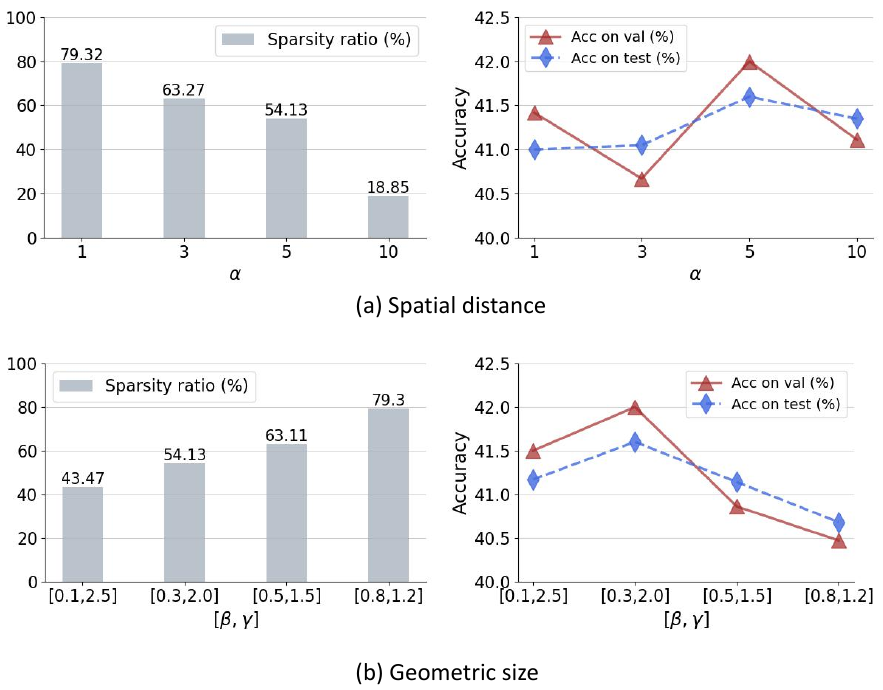}
\caption{Ablation studies of spatial factors in TSG.}
\vspace{-0.3cm}
\label{Figure 7}
\end{figure}

In this subsection, we conduct experiments with Rosetta-en OCR features on the TextVQA dataset to demonstrate the validity of our sparse method in Tables~\ref{tab:table4} $\sim$ Tables~\ref{tab:table10}, and \ref{tab:table-gat}, and to illustrate the role of heuristics in Tables~\ref{tab:table7}, \ref{tab:table8}, and Fig.~\ref{Figure 7}.
\subsubsection{\textbf{Sparsity Test}}
In this part, we discuss the graph sparsity in Tables~\ref{tab:table4} and \ref{tab:table5}. There are different sparsity settings. ``\textbf{w/ OTSG sparsity}" indicates that our approach just conducts the spatial sparsity of OTSG, while ``\textbf{w/o OTSG sparsity}" indicates that we just cancel the sparsity operation of OTSG. And the definitions of ``\textbf{w/ OSG sparsity}", ``\textbf{w/ TSG sparsity}", ``\textbf{w/o OSG sparsity}", and ``\textbf{w/o TSG sparsity}" are similar. For ``\textbf{w/o sparsity}", OTSG, OSG, and TSG are all performed on fully-connected graphs.

\begin{table}
\centering
\fontsize{10}{12}\selectfont
\caption{Statistics of sparsity ratio (SR) \textcolor{red}{\protect\footnotemark[4]} on TextVQA and ST-VQA test sets under different OCR systems.}
\vspace{-0.3cm}
\label{tab:table5}
\resizebox{0.48\textwidth}{!}{
\begin{tabular}{llccc}
\Xhline{1.4pt}
{\makecell[l]{\textbf{Dataset}}} & {\makecell[l]{\textbf{OCR System}}} & 
\makecell{\textbf{OTSG SR (\%)}} & \makecell{\textbf{OSG SR (\%)}}& \makecell{\textbf{TSG SR (\%)}}   \\ 
\Xhline{1.4pt}
\multirow{4}{*}{\makecell[c]{TextVQA }}  
& Rosetta-en  & 10.32 & 14.66 & 54.13  \\
& SBD-Trans  & 10.23 & 14.66 & 57.05  \\
& Google-OCR  & 10.00 & 14.66 & 55.63  \\
& Microsoft-OCR  & 9.85 & 14.66 & 53.68   \\
\hline
\multirow{4}{*}{\makecell[c]{ST-VQA }}  
& Rosetta-en  &  13.24 & 15.70 & 51.92     \\
& SBD-Trans   & 32.36 & 15.70 & 55.81 \\
& Google-OCR  & 13.26  & 15.70 & 55.51 \\
& Microsoft-OCR & 12.88 & 15.70 & 57.06  \\
\Xhline{1.4pt}
\end{tabular}}
\vspace{-0.3cm}
\end{table}
~\footnotetext[3]{Sparsity Ratio (\%) = $Avg$($\frac{N_{p}}{N_{I}}$), where $N_{p}$ and $N_{I}$ denote the number of pruned edges and the total number of edges per image respectively. }

\begin{table}[t]
\caption{Statistics of sparsity ratio (SR) of TSG on TextVQA and ST-VQA test sets under different data distributions of OCR token.}
\vspace{-0.3cm}
\label{tab:table10}
\fontsize{10}{12}\selectfont
\resizebox{0.48\textwidth}{!}{
\begin{tabular}{llcccc}
\Xhline{1.6pt}
\multirow{2}{*}{\textbf{Dataset}} & \multirow{2}{*}{\textbf{OCR System}} & \multicolumn{1}{l}{\textbf{TSG SR (\%)}} & \textbf{Data (\%)} & \multicolumn{1}{l}{\textbf{TSG SR (\%)}} & \textbf{Data (\%)} \\ \cline{3-6}  &  & \multicolumn{2}{c}{\textbf{OCR $\leqslant$ 20}} & \multicolumn{2}{c}{\textbf{OCR $\textgreater$ 20}}    \\ 
\Xhline{1.6pt}
\multirow{4}{*}{\makecell[c]{TextVQA}} & Rosetta-en & \multicolumn{1}{c}{56.29} & 86.57 & \multicolumn{1}{c}{40.91} & 13.43 \\  
& SBD-Trans & \multicolumn{1}{c}{60.35} & 78.18 & \multicolumn{1}{c}{45.55} & 21.82 \\  
& Google-OCR & \multicolumn{1}{c}{59.97} & 77.29 & \multicolumn{1}{c}{42.39} & 22.71 \\ 
& Microsoft-OCR & \multicolumn{1}{c}{57.79} & 73.18 & \multicolumn{1}{c}{42.98} & 26.82 \\ \hline
&&\multicolumn{2}{c}{\textbf{OCR $\leqslant$ 10}}    & \multicolumn{2}{c}{\textbf{OCR $\textgreater$ 10}}    \\ \hline 
\multirow{4}{*}{\makecell[c]{ST-VQA}} &   Rosetta-en & \multicolumn{1}{c}{55.26} & 80.79 & \multicolumn{1}{c}{38.30} & 19.21  \\
& SBD-Trans & \multicolumn{1}{c}{58.13} & 76.24 & \multicolumn{1}{c}{48.51} & 23.76 \\ 
& Google-OCR & \multicolumn{1}{c}{58.00} & 78.38 & \multicolumn{1}{c}{47.26} & 21.62 \\
& Microsoft-OCR & \multicolumn{1}{c}{65.95} & 70.40 & \multicolumn{1}{c}{44.25} & 29.60 \\ 
\Xhline{1.6pt}
\end{tabular}}
\vspace{-0.3cm}
\end{table}

\begin{table}[t]
\caption{Ablation studies of various IoUs in OSG on TextVQA dataset.}
\vspace{-0.3cm}
\label{tab:table7}
\centering
\fontsize{6}{7}\selectfont
\resizebox{0.48\textwidth}{!}{
\begin{tabular}{lcccc}
\Xhline{0.8pt}
\makecell[l]{\textbf{Method}} & \makecell[l]{\textbf{Threshold} \bm{$\epsilon$}} & \makecell[l]{\textbf{{SR (\%)}}}  & \makecell[c]{\textbf{Val Acc}} & \makecell[c]{\textbf{Test Acc}} \\
\Xhline{0.8pt}
\multirow{4}{*}{\makecell[c]{DIoU}} & $\epsilon$ = 0.01 & 24.59 & 41.15 &  40.93 \\
& $\epsilon$ = 0.1 & 19.71  & 41.51 & 41.02  \\
& \textbf{{$\epsilon$} = 0.3} & \textbf {14.66} & \textbf{42.00} & \textbf{41.60}  \\
& $\epsilon$ = 0.8 & 10.49  & 41.67 & 41.04  \\		
\hline
IoU & $\epsilon$ = 0.3 &15.33  & 41.47 & 40.66  \\
GIoU & $\epsilon$ = 0.3 &14.35 & 41.03 & 40.72  \\
CIoU & $\epsilon$ = 0.3 & 14.61 & 41.66 & 41.57  \\
\textbf{DIoU (Ours)} & \textbf{{$\epsilon$} = 0.3} & \textbf{14.66}  & \textbf{42.00} & \textbf{41.60}  \\
\Xhline{0.8pt}
\end{tabular}}
\vspace{-0.3cm}
\end{table}

\begin{table}[t]
\caption{{Ablation studies of threshold $\theta$ in OTSG and OSG on TextVQA dataset.}}
\vspace{-0.3cm}
\label{tab:table8}
\centering
\fontsize{5}{6}\selectfont
\resizebox{0.48\textwidth}{!}{
\begin{tabular}{lcccc}
\Xhline{0.8pt}
\makecell[l]{\textbf{Method}} & \makecell[l]{\textbf{Threshold $\theta$}} & \makecell[l]{\textbf{{SR (\%)}}}  & \makecell[c]{\textbf{Val Acc}} & \makecell[c]{\textbf{Test Acc}} \\
\Xhline{0.8pt}
\multirow{4}{*}{\makecell[c]{OTSG}} & $\theta$ = 0.3 & 42.24  & 41.35 & 40.98 \\
& $\theta$ = 0.4 & 22.78 & 41.56  &  41.21  \\
& $\theta$ = \textbf{0.5} & \textbf{10.32} & \textbf{42.00} & \textbf{41.60}  \\
& $\theta$ = 0.6 & 00.04  & 41.11  &  40.69  \\	
\hline
\multirow{4}{*}{\makecell[c]{OSG}} & $\theta$ = 0.3 & 46.15 &  41.10 &  41.03 \\
& $\theta$ = 0.4 &  26.89 &  41.51 &   41.13 \\
& $\theta$ = \textbf{0.5} & \textbf{14.66} & \textbf{42.00} & \textbf{41.60}  \\
& $\theta$ = 0.6 &  00.08 &  41.33 &  40.97  \\	
\Xhline{0.8pt}
\end{tabular}}
\vspace{-0.3cm}
\end{table}

\begin{table}[t]
\centering
\caption{{Ablation studies of soft solution based on GAT \cite{velivckovicgraph} technique with Rosetta-en OCR features on TextVQA dataset.}}
\vspace{-0.3cm}
\label{tab:table-gat}
\fontsize{8}{10}\selectfont
\resizebox{0.48\textwidth}{!}{
\begin{tabular}{lcc}
\Xhline{0.8pt}
\textbf{Method} & \textbf{\makecell[c]{Val Acc}} & \textbf{\makecell[c]{Test Acc}} \\
\Xhline{0.8pt}
SSGN-GAT  & 39.86 & 39.59  \\
SSGN-GAT-Soft Sparse (Hyperparameter) & 40.01 & 39.98  \\
SSGN-GAT-Soft Sparse (Median) & 39.99 & 39.87  \\
SSGN-GAT-Soft Sparse (Mean) & 39.94 & 39.73  \\
SSGN-GAT-Spatial Sparse & 40.70 & 40.35 \\
\hline
\textbf{SSGN (Ours)} & \textbf{42.00} & \textbf{41.60} \\
\Xhline{0.8pt}
\end{tabular}}
\end{table}

As shown in Table~\ref{tab:table4}, under only one sub-graph sparsity, the effect of \textbf{w/ OTSG sparsity} is most noticeable due to the importance of object-OCR token correlation. If a local region in the image is recognized by both objects and OCR tokens, it does deserve more attention. Building these relations with object-OCR co-occurrence are instructive for predicting answers. Besides, the performances are very close in the graph variants of two sub-graph sparsity, where \textbf{w/o TSG sparsity} performs marginally better than others. In fact, due to the characteristics of object detectors and OCR systems, dense object-object relations with large overlapping regions widely exist, while overlapping OCR {tokens} are much less frequent. Anyway, \textbf{SSGN} with the complete sparsity setting achieves the best results. Furthermore, the sparsity ratios of OTSG, OSG, and TSG are reported in Table~\ref{tab:table5}, around 10\%, 15\%, and 55\%, respectively. These sparsity ratios are roughly stable, except for the case of 32.36\% OTSG under SBD-Trans features on ST-VQA. We are surprised to find that this is actually caused by the SBD-Trans OCR system's mislabeling of object bounding boxes on the coco-text subset of the ST-VQA dataset. Among these, TSG has the greatest sparsity.

Back to the statistics of OCR tokens in Table~\ref{tab:table9}, various OCR systems output different but competitive token numbers with each other. Here, we discuss the effect of OCR systems on the sparsity rate in Table~\ref{tab:table10}. We choose the \emph{Mean} number of tokens per image as the cut-off point to observe the sparsity ratio of TSG, \ie, 20 for the TextVQA dataset and 10 for the ST-VQA dataset. As shown in Table~\ref{tab:table10}, the TSG sparsity ranges from 55\% to 66\% for OCR$\leqslant$20/OCR$\leqslant$10 on the TextVQA / ST-VQA, while is about 45\% for OCR$>$20/OCR$>$10, respectively. In conclusion, the sparsity patterns of the two datasets are similar and we consider the adaptive sparsity strategy in our approach to be stable and acceptable. 

\subsubsection{\textbf{Impact of various IoUs}} 
IoU is widely used for object detection. In this part, we discuss its effect in the OSG graph (Eq.~\ref{eq:eq11}). There are many variants of IoU, including IoU\cite{everingham2010pascal}, GIoU \cite{rezatofighi2019generalized}, CIoU \cite{zheng2020distance} and DIoU \cite{zheng2020distance}. {IoU} \cite{everingham2010pascal} is a basic term that considers the overlap area of two object bounding boxes. With the basic IoU, {GIoU (Generalized IoU)} \cite{rezatofighi2019generalized} considers the relative direction, {DIoU} \cite{zheng2020distance} (Distance-IoU) adds the center measurement, and {CIoU} (Complete-DIoU) \cite{zheng2020distance} adds the length and width measurements of bounding box. As shown in Table~\ref{tab:table7}, \textbf{DIoU} considering overlap area and center distance performs the best at \textbf{$\epsilon$} = 0.3. Using only overlap area is not sufficient (\textbf{IoU} drops by 0.53\%), and considering direction and angle is not appropriate for edge modeling in the TextVQA task (\textbf{GIoU} drops by 0.97 \%). \textbf{CIoU} performs well (down 0.34\%), but the length and width measurements are not effective to \textbf{DIoU} (considering center distance) in eliminating redundant spatial relations.

\subsubsection{\textbf{Impact of Spatial Factor}} 
{We analyze the role of spatial factors $\theta$ in graphs OTSG and OSG, and test $\{\alpha, \gamma\}$ in the graph TSG. In OTSG, $\theta$ is used to constrain the spatial distance or their overlap area of object-token pair, while $\theta$ is taken to constrain the spatial distance of object-object pair in OSG. As shown in Table~\ref{tab:table8}, the greater the value of $\theta$ is, the fewer edges are pruned. The results show that the spatial distance is too close or too far to be suitable for relation inference. $\theta$ = 0.5 is the optimal setup for OTSG and OSG.} We further test the spatial factors of distance and geometric size in the graph TSG. As shown in Fig.~\ref{Figure 7}, the larger $\alpha$ is, the {fewer} edges are pruned. We set $\alpha$ = 5 for the best performance. We consider that there is a balance between the sparsity ratio and distance. For the token's geometric size [$\beta$, $\gamma$], the setting of $[0.3, 2.0]$ achieves the best performance. It seems that the broader ranges $[0.1, 2.5]$ and $[0.3, 2.0]$ filter out the relations useful for answer prediction better than $[0.5, 1.5]$ and $[0.8, 1.2]$.

\begin{figure*}
\centering
\includegraphics[width=\textwidth]{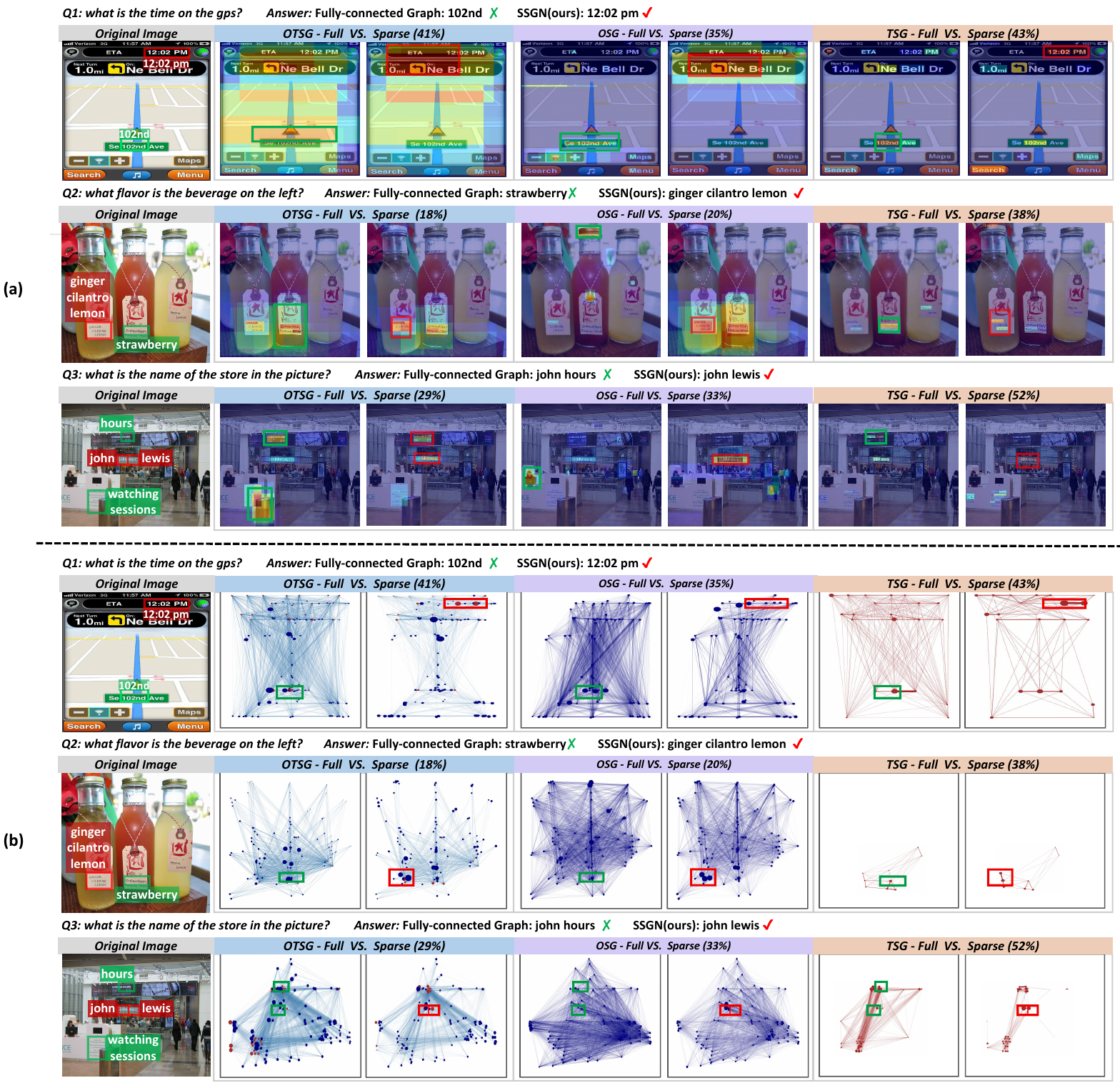}
\caption{Instantiation of fully connected and sparse graphs for OTSG, OSG, and TSG. The number in each parenthesis indicates the sparsity ratio. We show in (a) the visualization of nodes (objects and OCR tokens) overlapping in the image and in (b) the visualization of edges (relations). The node size reflects the sum of the connected edge weights and each edge weight is output by the message transition matrices ${\mathcal{A}^{vt}}$, ${\mathcal{A}^{v}}$, and ${\mathcal{A}^{t}}$	of each graph. By comparison, the sparse graph is more resistant to the interference of redundant relations and can generate accurate answers.}
\vspace{-0.3cm}
\label{Figure 4}
\end{figure*}

\begin{figure*}
\centering
\includegraphics[width=\textwidth]{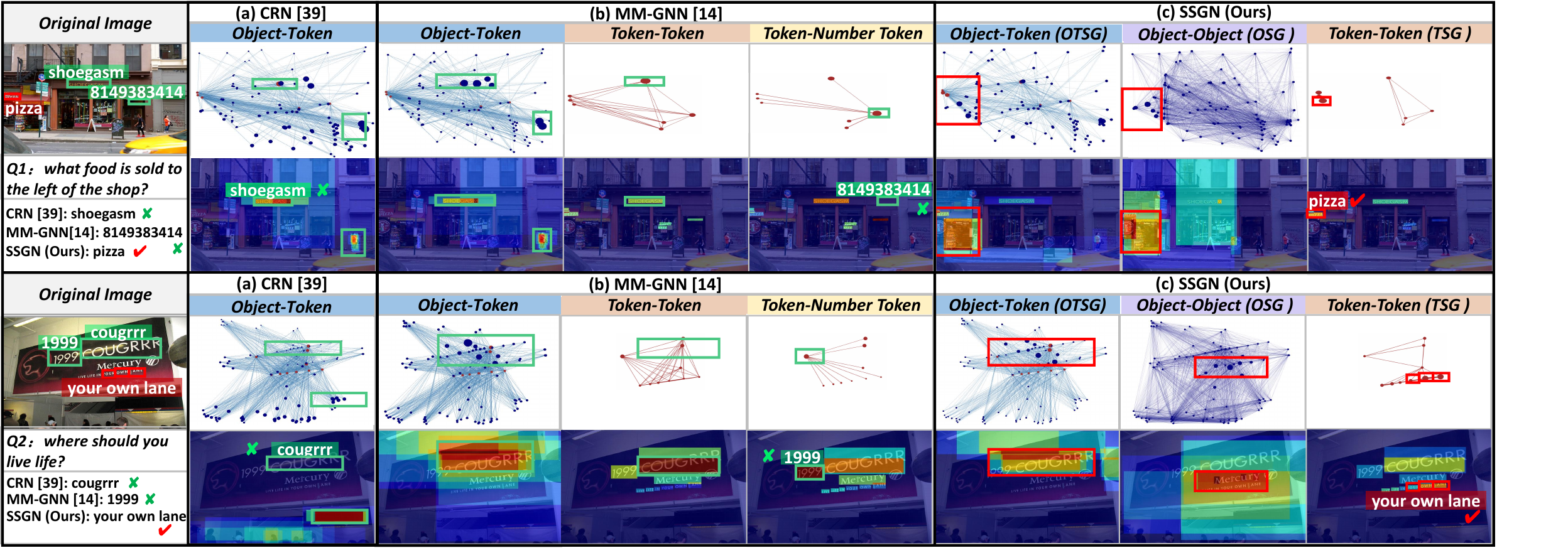}
\caption{{
Two visualization examples of our method compared with existing CRN \cite{liu2020cascade} and MM-GNN \cite{gao2020multi}. The results show that through effective spatial pruning, our method SSGN performs accurate relation inference between object-token, object-object, and token-token for answer prediction.
}}
\label{Fig_vs_graphs}
\vspace{-0.1cm}
\end{figure*}

{\subsubsection{\textbf{Soft Relation Pruning}} Here, we consider a flexible soft relational solution referring to GAT \cite{velivckovicgraph} technique. We replace the relational learning mode in the OTSG, OSG, and TSG graphs with soft weight learning in GAT \cite{velivckovicgraph}. As shown in Table~\ref{tab:table-gat}, the variants for ablation experiments are as follows: 1) \textbf{SSGN-GAT} is a fully-connected graph model that uses soft weights to combine full relations; 2) based on it, we test three soft pruning methods, \ie, \textbf{SSGN-GAT-Soft Sparse (Hyperparameter)}, \textbf{SSGN-GAT-Soft Sparse (Median)}, and \textbf{SSGN-GAT-Soft Sparse (Mean)}. The three sparse graph models respectively use an empirical \emph{hyperparameter} (0.01), \emph{median} and \emph{mean} of edge weights to adaptively select sparse relations with different sparse criteria;
3) different from soft sparse solution, \textbf{SSGN-GAT-Spatial Sparse} is a sparse graph model that uses spatial conditions proposed in this work. Among these sparse graph variants, the baseline \textbf{SSGN-GAT} performs the worst. Compared with \textbf{SSGN-GAT}, the performances of all SSGN-GAT-Soft variants are slightly improved; for examples, \textbf{SSGN-GAT-Soft Sparse (Hyperparameter)} improves 0.15\% and 0.39\% on the val and test sets, respectively. Compared with \textbf{SSGN-GAT}, \textbf{SSGN-GAT-Spatial Sparse} with our spatial constraints increases 0.84\% and 0.76\% on the val and test sets, respectively. It proves that our spatial pruning method with sophisticated spatial conditions is effective and conducive to answer reasoning. Anyway, the proposed method \textbf{SSGN (Ours)} performs the best.
}

\subsection{Visualization Analysis}
\label{Visualization Analysis}
To demonstrate the interpretability of our method, we visualize some examples below. 

\begin{figure}[t]
\centering
\includegraphics[width=\columnwidth]{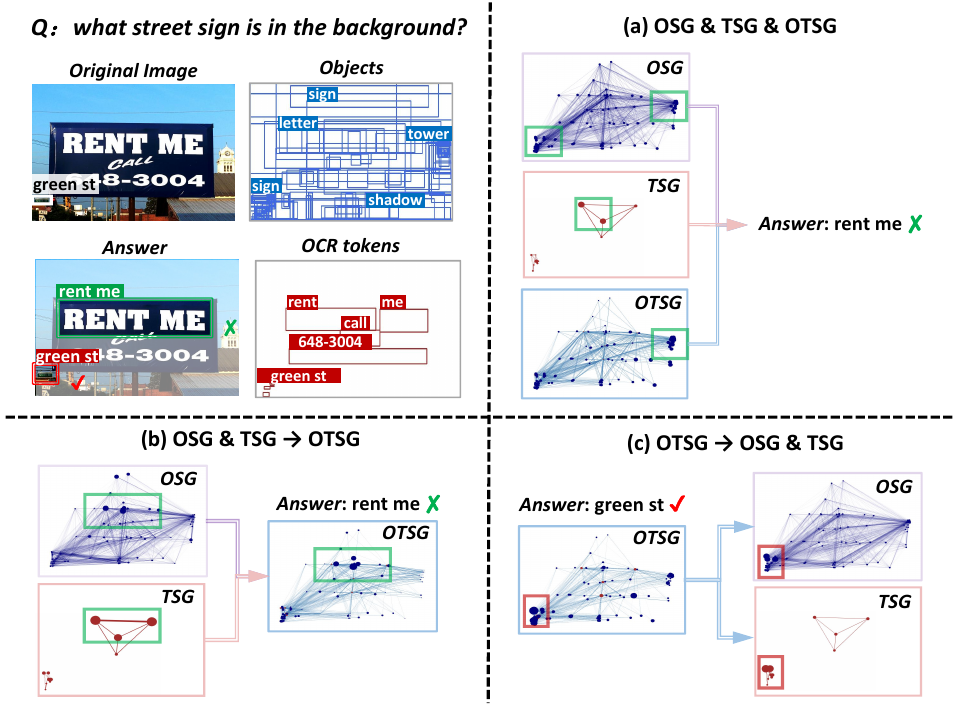}
\caption{Visualization of hierarchical graph structures for answer prediction. We discuss three solutions, namely {OSG\&TSG\&OTSG}, {OSG\&TSG$\rightarrow$OTSG}, and {OTSG$\rightarrow$OSG\&TSG}. By observing, in our method {OTSG$\rightarrow$OSG\&TSG}, the implementation of OTSG can first effectively discover the critical relations through the nearby object and OCR token co-occurrences, and then the parallel learning of OSG and TSG examines the critical relations under each space of objects and OCR tokens. } 
\label{Figure 10}
\vspace{-0.5cm}
\end{figure}

\begin{figure}[t]
\centering
\includegraphics[width=\columnwidth]{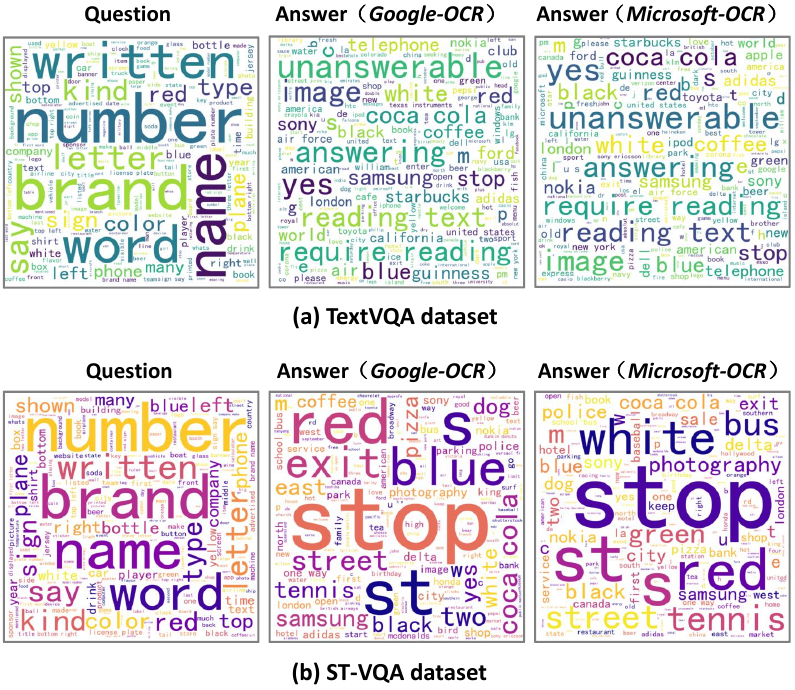}
\caption{Question and answer word clouds for TextVQA and ST-VQA datasets. We visualize the generated answers with Google-OCR and Microsoft-OCR as examples. For both questions and answers, stop words, \eg, "the", "is", "which", "what", "at", "on", \etc, are removed from the statistics.}
\label{Figure 6}
\vspace{-0.3cm}
\end{figure}

\subsubsection{\textbf{Graph Inference}}
As shown in Fig.~\ref{Figure 4} {(a)}, we exhibit the graph learning process in the fully-connected graph and the sparse graph settings. The highly-responsive visual regions are quite different in the two graph settings. For the questions \textbf{Q1}$\sim$\textbf{Q3}, in fully-connected graphs, the misleading answers are ``102nd", ``strawberry" and ``john hour", while in sparse graphs the correct answers are ``12:02 pm”, ``ginger cilantro lemon", and ``john lewis". A remarkable observation is the existence of redundant relations in fully-connected graphs. For example in \textbf{Q1}, at the bottom of the image, {the green road sign is recognized as a ``sign" by the object detection model, which is covered by 11 different-sized bounding boxes that are considered as 11 objects.} These close-by objects enhance the unnecessary visual relations between them. {And the OCR token ``Search” is far from the token ``12:02 pm”}, while the relations between these two OCR tokens is no longer semantically needed. In the sparse graph, we cut off the redundant connections and make the inference of the answer more explicit. Fig.~\ref{Figure 4} gives an illustrative explanation of the effectiveness of sparse graph learning in this work.

\subsubsection{\textbf{Spatial Sparsity Analysis}}
As shown in Fig.~\ref{Figure 4} {(b)}, in terms of the sparsity of OCR tokens, in \textbf{Q3}, the three tokens ``john", ``lewis" and ``hours" are tightly connected in the fully-connected TSG graph. In fact, the two tokens ``john" and ``lewis" (the correct answer) are spatially close to each other but relatively distant from the token ``hours". After relation pruning in the graph, the semantic difference between ``hours" and ``john lewis" is more explicit for the model. How about the other sparsity? Taking \textbf{Q1} ``what is the time on the gaps" as an example, it is hard to distinguish between the numeric tokens ``102nd" and ``12:02 pm". In the fully-connected graph setting, the effect of ``102nd" is enhanced by the redundant object-object and object-OCR token relations. After spatial pruning, ``102nd" and ``12:02 pm" can be evaluated fairly, and the correct answer ``12:02 pm" is thus output.

{\subsubsection{\textbf{Graph Model Comparison}}
Here, we display two examples to compare our model with two existing fully-connected graph methods \textbf{CRN} \cite{liu2020cascade} and \textbf{MM-GNN} \cite{gao2020multi}. As shown in the Fig.~\ref{Fig_vs_graphs}, it can be found that both \textbf{CRN} \cite{liu2020cascade} and \textbf{MM-GNN} \cite{gao2020multi} are confused by the full relations and give wrong answers, while the proposed method performs well attributing to the sparse relation learning.
As shown in \textbf{Q1 (a)} and \textbf{Q1 (b)}, the redundant relation between object ``woman" and token ``shoegasm" interferes with the  answer reasoning of \textbf{CRN} \cite{liu2020cascade} to the wrong answer ``shoegasm", while the dense relations between token ``shoegasm", digital token ``8149383414" and the other tokens mislead the model \textbf{MM-GNN} \cite{gao2020multi}, resulting the wrong answer ``8149383414". In \textbf{Q1 (c)}, \textbf{SSGN (Our)} cuts off the relations between spatially distant object-token pairs and reduces the redundant associations between repeated objects and disconnected tokens by leveraging the customized spatial criteria. For example, in our TSG (token-token) graph, the token ``pizza" has no connection to the tokens ``shoegasm" and ``8149383414". Finally, our model reasons out the correct answer ``pizza". In addition, as shown in \textbf{Q2 (c)}, a similar conclusion can be drawn in \textbf{Q2 (c)}. The valuable and effective semantic associations are helpful for answer prediction rather than fully semantic associations established in the graphs.
}

\subsubsection{\textbf{Hierarchical Graph Structure}}
\label{Different Hierarchical Graph Inference}
Here, we discuss the hierarchical graph structure for answer prediction. We visualize an example in Fig.~\ref{Figure 10}. The proposed SSGN method is carried out with the parallel structure \textbf{OSG\&TSG\&OTSG}, as well as the cascading structure of the order  \textbf{OSG\&TSG$\rightarrow$OTSG} and its reverse order \textbf{OTSG$\rightarrow$OSG\&TSG}, respectively. To answer the question ``what street sign is in the background?" In \textbf{OSG\&TSG\&OTSG}, each sub-graph focuses on different entities in each visual space, such as object nodes ``tower" and ``sign" in OSG, OCR token nodes ``rent me" and ``call" in TSG and object node ``tower" in OTSG. Unlike the discovery in \textbf{OSG\&TSG\&OTSG}, \textbf{OSG\&TSG$\rightarrow$OTSG} and \textbf{OTSG$\rightarrow$OSG\&TSG} consistently focus on ``sign" in the image. But \textbf{OSG\&TSG$\rightarrow$OTSG} performs the graph learning in separate object and OCR token spaces in parallel first, resulting in a wrong focus on the foreground ``sign". In contrast, our \textbf{OTSG$\rightarrow$OSG\&TSG} outputs the correct answer. By first implementing the object-OCR token correlation, our approach focuses directly on the background ``sign".

\subsubsection{\textbf{Word Cloud Analysis}} Here, we use the word clouds to visualize the high-frequency words in the questions and answers. As shown in Fig.~\ref{Figure 6}, the questions in both TextVQA and ST-VQA datasets pay consistent attention to words ``\emph{name}", ``\emph{number}", ``\emph{word}", ``\emph{brand}", and ``\emph{written}". As for the answers, due to the difficulty of questions, ``\emph{unanswerable}" occurs with a remarkable frequency in TextVQA dataset. And the road sign word "\emph{stop}" appears most frequently in the ST-VQA answers. It is also interesting to note that the ST-VQA prefers to ask questions about color accompanied with the answers ``\emph{blue}", ``\emph{red}", and ``\emph{white}".

\section{Conclusion}
\label{subsec:Conclusion}
In this paper, we propose a sparse spatial graph network (SSGN) for TextVQA, which focuses on edge pruning in graph learning. We investigate a depth study of graph sparsity from spatial factors, such as DIoU, distance, geometric size, and overlap area. We strive to prune redundant or useless relations. Extensive experiments are conducted on TextVQA and ST-VQA datasets under different OCR systems to validate the effectiveness of SSGN and to show interpretable visualization results. 

{
\bibliographystyle{ieee_fullname}
\bibliography{ref}

\begin{thebibliography}{10}\itemsep=-1pt

\bibitem{almazan2014word}
Jon Almaz{\'a}n, Albert Gordo, Alicia Forn{\'e}s, and Ernest Valveny.
\newblock Word spotting and recognition with embedded attributes.
\newblock {\em IEEE TPAMI}, pages 2552--2566, 2014.

\bibitem{Baek2019WhatIW}
Jeonghun Baek, Geewook Kim, Junyeop Lee, Sungrae Park, Dongyoon Han, Sangdoo
  Yun, Seong~Joon Oh, and Hwalsuk Lee.
\newblock What is wrong with scene text recognition model comparisons? dataset
  and model analysis.
\newblock In {\em ICCV}, pages 4714--4722, 2019.

\bibitem{biten2019scene}
Ali~Furkan Biten, Ruben Tito, Andres Mafla, Lluis Gomez, Marcal Rusinol, CV
  Jawahar, Ernest Valveny, and Dimosthenis Karatzas.
\newblock Scene text visual question answering.
\newblock In {\em ICCV}, pages 4290--4300, 2019.

\bibitem{Biten2019ICDAR2C}
Ali~Furkan Biten, Rub{\`e}n~P{\'e}rez Tito, Andr{\'e}s Mafla, Llu{\'i}s
  G{\'o}mez, Marçal Rusi{\~n}ol, Minesh Mathew, C.~V. Jawahar, Ernest Valveny,
  and Dimosthenis Karatzas.
\newblock {{ICDAR}} 2019 competition on scene text visual question answering.
\newblock In {\em ICDAR}, pages 1563--1570, 2019.

\bibitem{bojanowski2017enriching}
Piotr Bojanowski, Edouard Grave, Armand Joulin, and Tomas Mikolov.
\newblock Enriching word vectors with subword information.
\newblock {\em TACL}, pages 135--146, 2017.

\bibitem{borisyuk2018rosetta}
Fedor Borisyuk, Albert Gordo, and Viswanath Sivakumar.
\newblock Rosetta: Large scale system for text detection and recognition in
  images.
\newblock In {\em SIGKDD}, pages 71--79, 2018.

\bibitem{chen2021gog}
Feilong Chen, Xiuyi Chen, Fandong Meng, Peng Li, and Jie Zhou.
\newblock Gog: Relation-aware graph-over-graph network for visual dialog.
\newblock In {\em ACL}, pages 230--243, 2021.

\bibitem{chen2020monopair}
Yongjian Chen, Lei Tai, Kai Sun, and Mingyang Li.
\newblock Monopair: Monocular 3d object detection using pairwise spatial
  relationships.
\newblock In {\em CVPR}, pages 12093--12102, 2020.

\bibitem{deng2009imagenet}
Jia Deng, Wei Dong, Richard Socher, Li-Jia Li, Kai Li, and Li Fei-Fei.
\newblock Imagenet: A large-scale hierarchical image database.
\newblock In {\em CVPR}, pages 248--255, 2009.

\bibitem{Devlin2019BERTPO}
Jacob Devlin, MingWei Chang, Kenton Lee, and Kristina Toutanova.
\newblock Bert: Pre-training of deep bidirectional transformers for language
  understanding.
\newblock In {\em NAACL}, page 4171–4186, 2019.

\bibitem{everingham2010pascal}
Mark Everingham, Luc Van~Gool, Christopher~KI Williams, John Winn, and Andrew
  Zisserman.
\newblock The pascal visual object classes (voc) challenge.
\newblock {\em IJCV}, pages 303--338, 2010.

\bibitem{Fang2021ReadLH}
Shancheng Fang, Hongtao Xie, Yuxin Wang, Zhendong Mao, and Yongdong Zhang.
\newblock Read like humans: Autonomous, bidirectional and iterative language
  modeling for scene text recognition.
\newblock In {\em CVPR}, pages 7094--7103, 2021.

\bibitem{gao2020structured}
Chenyu Gao, Qi Zhu, Peng Wang, Hui Li, Yuliang Liu, Anton Van~den Hengel, and
  Qi Wu.
\newblock Structured multimodal attentions for textvqa.
\newblock {\em IEEE TPAMI}, 2021.

\bibitem{gao2020multi}
Difei Gao, Ke Li, Ruiping Wang, Shiguang Shan, and Xilin Chen.
\newblock Multi-modal graph neural network for joint reasoning on vision and
  scene text.
\newblock In {\em CVPR}, pages 12746--12756, 2020.

\bibitem{Gu2021GraphBasedMN}
Mao Gu, Zhou Zhao, Weike Jin, Richang Hong, and Fei Wu.
\newblock Graph-based multi-interaction network for video question answering.
\newblock {\em IEEE TIP}, pages 2758--2770, 2021.

\bibitem{guo2019dadnet}
Dan Guo, Kun Li, Zheng-Jun Zha, and Meng Wang.
\newblock Dadnet: Dilated-attention-deformable convnet for crowd counting.
\newblock In {\em Proceedings of the 27th ACM international conference on
  multimedia}, pages 1823--1832, 2019.

\bibitem{guo2019dense}
Dan Guo, Shuo Wang, Qi Tian, and Meng Wang.
\newblock Dense temporal convolution network for sign language translation.
\newblock In {\em IJCAI}, pages 744--750, 2019.

\bibitem{guo2018hierarchical}
Dan Guo, Wengang Zhou, Houqiang Li, and Meng Wang.
\newblock Hierarchical lstm for sign language translation.
\newblock In {\em AAAI}, 2018.

\bibitem{Guo2021ReAttentionFV}
Wenya Guo, Ying Zhang, Jufeng Yang, and Xiaojie Yuan.
\newblock Re-attention for visual question answering.
\newblock {\em IEEE TIP}, pages 6730--6743, 2021.

\bibitem{gurari2018vizwiz}
Danna Gurari, Qing Li, Abigale~J Stangl, Anhong Guo, Chi Lin, Kristen Grauman,
  Jiebo Luo, and Jeffrey~P Bigham.
\newblock Vizwiz grand challenge: Answering visual questions from blind people.
\newblock In {\em CVPR}, pages 3608--3617, 2018.

\bibitem{han2020finding}
Wei Han, Hantao Huang, and Tao Han.
\newblock Finding the evidence: Localization-aware answer prediction for text
  visual question answering.
\newblock In {\em COLING}, pages 3118--3131, 2020.

\bibitem{hu2020iterative}
Ronghang Hu, Amanpreet Singh, Trevor Darrell, and Marcus Rohrbach.
\newblock Iterative answer prediction with pointer-augmented multimodal
  transformers for textvqa.
\newblock In {\em CVPR}, pages 9992--10002, 2020.

\bibitem{huang2020aligned}
Qingbao Huang, Jielong Wei, Yi Cai, Changmeng Zheng, Junying Chen, Hofung
  Leung, and Qing Li.
\newblock Aligned dual channel graph convolutional network for visual question
  answering.
\newblock In {\em ACL}, pages 7166--7176, 2020.

\bibitem{Hudson2019GQAAN}
Drew~A. Hudson and Christopher~D. Manning.
\newblock Gqa: A new dataset for real-world visual reasoning and compositional
  question answering.
\newblock In {\em CVPR}, pages 6693--6702, 2019.

\bibitem{inayoshi2020bounding}
Sho Inayoshi, Keita Otani, Antonio Tejero-de Pablos, and Tatsuya Harada.
\newblock Bounding-box channels for visual relationship detection.
\newblock In {\em ECCV}, pages 682--697, 2020.

\bibitem{Ji2020SpatioTemporalMA}
Junzhong Ji, Cheng Xu, Xiaodan Zhang, Boyue Wang, and Xinhang Song.
\newblock Spatio-temporal memory attention for image captioning.
\newblock {\em IEEE TIP}, pages 7615--7628, 2020.

\bibitem{jiang2018pythia}
Yu Jiang, Vivek Natarajan, Xinlei Chen, Marcus Rohrbach, Dhruv Batra, and Devi
  Parikh.
\newblock Pythia v0. 1: the winning entry to the vqa challenge 2018.
\newblock {\em arXiv preprint arXiv:1807.09956}, 2018.

\bibitem{jin2021ruart}
ZanXia Jin, Heran Wu, Chun Yang, Fang Zhou, Jingyan Qin, Lei Xiao, and XuCheng
  Yin.
\newblock Ruart: A novel text-centered solution for text-based visual question
  answering.
\newblock {\em IEEE TMM}, pages 1--1, 2021.

\bibitem{Kafle2018DVQAUD}
Kushal Kafle, Scott~D. Cohen, Brian~L. Price, and Christopher Kanan.
\newblock Dvqa: Understanding data visualizations via question answering.
\newblock In {\em CVPR}, pages 5648--5656, 2018.

\bibitem{kant2020spatially}
Yash Kant, Dhruv Batra, Peter Anderson, Alexander Schwing, Devi Parikh, Jiasen
  Lu, and Harsh Agrawal.
\newblock Spatially aware multimodal transformers for textvqa.
\newblock In {\em ECCV}, pages 715--732, 2020.

\bibitem{karatzas2015icdar}
Dimosthenis Karatzas, Lluis Gomez-Bigorda, Anguelos Nicolaou, Suman Ghosh,
  Andrew Bagdanov, Masakazu Iwamura, Jiri Matas, Lukas Neumann,
  Vijay~Ramaseshan Chandrasekhar, Shijian Lu, et~al.
\newblock {{ICDAR}} 2015 competition on robust reading.
\newblock In {\em ICDAR}, pages 1156--1160, 2015.

\bibitem{karatzas2013icdar}
Dimosthenis Karatzas, Faisal Shafait, Seiichi Uchida, Masakazu Iwamura,
  Lluis~Gomez i Bigorda, Sergi~Robles Mestre, Joan Mas, David~Fernandez Mota,
  Jon~Almazan Almazan, and Lluis~Pere De~Las~Heras.
\newblock {{ICDAR}} 2013 robust reading competition.
\newblock In {\em ICDAR}, pages 1484--1493, 2013.

\bibitem{kim2020hypergraph}
EunSol Kim, Woo~Young Kang, KyoungWoon On, YuJung Heo, and ByoungTak Zhang.
\newblock Hypergraph attention networks for multimodal learning.
\newblock In {\em CVPR}, pages 14581--14590, 2020.

\bibitem{krasin2017openimages}
Ivan Krasin, Tom Duerig, Neil Alldrin, Vittorio Ferrari, Sami AbuElHaija, Alina
  Kuznetsova, Hassan Rom, Jasper Uijlings, Stefan Popov, Andreas Veit, et~al.
\newblock Openimages: A public dataset for large-scale multi-label and
  multi-class image classification.
\newblock {\em Dataset available from https://github. com/openimages}, 2017.

\bibitem{krishna2017visual}
Ranjay Krishna, Yuke Zhu, Oliver Groth, Justin Johnson, Kenji Hata, Joshua
  Kravitz, Stephanie Chen, Yannis Kalantidis, Li-Jia Li, David~A Shamma, et~al.
\newblock Visual genome: Connecting language and vision using crowdsourced
  dense image annotations.
\newblock {\em IJCV}, pages 32--73, 2017.

\bibitem{levenshtein1966binary}
Vladimir~I Levenshtein et~al.
\newblock Binary codes capable of correcting deletions, insertions, and
  reversals.
\newblock In {\em Soviet physics doklady}, pages 707--710, 1966.

\bibitem{li2019relation}
Linjie Li, Zhe Gan, Yu Cheng, and Jingjing Liu.
\newblock Relation-aware graph attention network for visual question answering.
\newblock In {\em ICCV}, pages 10313--10322, 2019.

\bibitem{Liao2022ProgressiveLV}
Yue Liao, Aixi Zhang, Zhiyuan Chen, Tianrui Hui, and Si Liu.
\newblock Progressive language-customized visual feature learning for one-stage
  visual grounding.
\newblock {\em IEEE TIP}, pages 4266--4277, 2022.

\bibitem{liu2020cascade}
Fen Liu, Guanghui Xu, Qi Wu, Qing Du, Wei Jia, and Mingkui Tan.
\newblock Cascade reasoning network for text-based visual question answering.
\newblock In {\em ACM MM}, pages 4060--4069, 2020.

\bibitem{Liu2019OmnidirectionalST}
Yuliang Liu, Sheng Zhang, Lianwen Jin, Lele Xie, Y. Wu, and Zhepeng Wang.
\newblock Omnidirectional scene text detection with sequential-free box
  discretization.
\newblock In {\em IJCAI}, pages 3052--3058, 2019.

\bibitem{Lu2021LocalizeGA}
XiaoPeng Lu, Zhenhua Fan, Yansen Wang, Jean Oh, and Carolyn~Penstein Ros{\'e}.
\newblock Localize, group, and select: Boosting text-vqa by scene text
  modeling.
\newblock In {\em ICCV workshop}, pages 2631--2639, 2021.

\bibitem{mishra2013image}
Anand Mishra, Karteek Alahari, and CV Jawahar.
\newblock Image retrieval using textual cues.
\newblock In {\em ICCV}, pages 3040--3047, 2013.

\bibitem{Ren2015FasterRT}
Shaoqing Ren, Kaiming He, Ross~B. Girshick, and Jian Sun.
\newblock Faster r-cnn: Towards real-time object detection with region proposal
  networks.
\newblock {\em IEEE TPAMI}, pages 1137--1149, 2015.

\bibitem{rezatofighi2019generalized}
Hamid Rezatofighi, Nathan Tsoi, JunYoung Gwak, Amir Sadeghian, Ian Reid, and
  Silvio Savarese.
\newblock Generalized intersection over union: A metric and a loss for bounding
  box regression.
\newblock In {\em CVPR}, pages 658--666, 2019.

\bibitem{sidorov2020textcaps}
Oleksii Sidorov, Ronghang Hu, Marcus Rohrbach, and Amanpreet Singh.
\newblock Textcaps: a dataset for image captioning with reading comprehension.
\newblock In {\em ECCV}, pages 742--758. Springer, 2020.

\bibitem{singh2019towards}
Amanpreet Singh, Vivek Natarajan, Meet Shah, Yu Jiang, Xinlei Chen, Dhruv
  Batra, Devi Parikh, and Marcus Rohrbach.
\newblock Towards vqa models that can read.
\newblock In {\em CVPR}, pages 8317--8326, 2019.

\bibitem{vaswani2017attention}
Ashish Vaswani, Noam Shazeer, Niki Parmar, Jakob Uszkoreit, Llion Jones,
  Aidan~N Gomez, {\L}ukasz Kaiser, and Illia Polosukhin.
\newblock Attention is all you need.
\newblock In {\em NeurIPS}, pages 5998--6008, 2017.

\bibitem{veit2016coco}
Andreas Veit, Tomas Matera, Lukas Neumann, Jiri Matas, and Serge Belongie.
\newblock Coco-text: Dataset and benchmark for text detection and recognition
  in natural images.
\newblock {\em arXiv preprint arXiv:1601.07140}, 2016.

\bibitem{velivckovicgraph}
Petar Veli{\v{c}}kovi{\'c}, Guillem Cucurull, Arantxa Casanova, Adriana Romero,
  Pietro Li{\`o}, and Yoshua Bengio.
\newblock Graph attention networks.
\newblock In {\em ICLR}, 2018.

\bibitem{wang2020multimodal}
Jing Wang, Jinhui Tang, and Jiebo Luo.
\newblock Multimodal attention with image text spatial relationship for
  ocr-based image captioning.
\newblock In {\em ACM MM}, pages 4337--4345, 2020.

\bibitem{wang2021improving}
Jing Wang, Jinhui Tang, Mingkun Yang, Xiang Bai, and Jiebo Luo.
\newblock Improving ocr-based image captioning by incorporating geometrical
  relationship.
\newblock In {\em CVPR}, pages 1306--1315, 2021.

\bibitem{wang2020general}
Xinyu Wang, Yuliang Liu, Chunhua Shen, Chun~Chet Ng, Canjie Luo, Lianwen Jin,
  Chee~Seng Chan, Anton van~den Hengel, and Liangwei Wang.
\newblock On the general value of evidence, and bilingual scene-text visual
  question answering.
\newblock In {\em CVPR}, pages 10126--10135, 2020.

\bibitem{wang2022progressive}
Yang Wang, Jinjia Peng, Huibing Wang, and Meng Wang.
\newblock Progressive learning with multi-scale attention network for
  cross-domain vehicle re-identification.
\newblock {\em Science China Information Sciences}, page 160103, 2022.

\bibitem{yang2020relationship}
Sibei Yang, Guanbin Li, and Yizhou Yu.
\newblock Relationship-embedded representation learning for grounding referring
  expressions.
\newblock {\em IEEE TPAMI}, pages 2765--2779, 2020.

\bibitem{yang2021tap}
Zhengyuan Yang, Yijuan Lu, Jianfeng Wang, Xi Yin, Dinei Florencio, Lijuan Wang,
  Cha Zhang, Lei Zhang, and Jiebo Luo.
\newblock Tap: Text-aware pre-training for text-vqa and text-caption.
\newblock In {\em CVPR}, pages 8751--8761, 2021.

\bibitem{yao2018exploring}
Ting Yao, Yingwei Pan, Yehao Li, and Tao Mei.
\newblock Exploring visual relationship for image captioning.
\newblock In {\em ECCV}, pages 684--699, 2018.

\bibitem{zeng2021beyond}
Gangyan Zeng, Yuan Zhang, Yu Zhou, and Xiaomeng Yang.
\newblock Beyond ocr+ vqa: involving ocr into the flow for robust and accurate
  textvqa.
\newblock In {\em ACM MM}, pages 376--385, 2021.

\bibitem{zhang2021position}
Xuanyu Zhang and Qing Yang.
\newblock Position-augmented transformers with entity-aligned mesh for textvqa.
\newblock In {\em ACM MM}, pages 2519--2528, 2021.

\bibitem{zheng2020distance}
Zhaohui Zheng, Ping Wang, Wei Liu, Jinze Li, Rongguang Ye, and Dongwei Ren.
\newblock Distance-iou loss: Faster and better learning for bounding box
  regression.
\newblock In {\em AAAI}, pages 12993--13000, 2020.

\bibitem{Zhu2021SimpleIN}
Qi Zhu, Chenyu Gao, P. Wang, and Qi Wu.
\newblock Simple is not easy: A simple strong baseline for textvqa and
  textcaps.
\newblock In {\em AAAI}, pages 3608--3615, 2021.

\bibitem{Zhu2020MuckoMC}
Zihao Zhu, J. Yu, Yujing Wang, Yajing Sun, Yue Hu, and Qi Wu.
\newblock Mucko: Multi-layer cross-modal knowledge reasoning for fact-based
  visual question answering.
\newblock In {\em IJCAI}, 2020.

\end{thebibliography}
}

\begin{IEEEbiography}[{\includegraphics[width=1in,height=1.25in,clip,keepaspectratio]{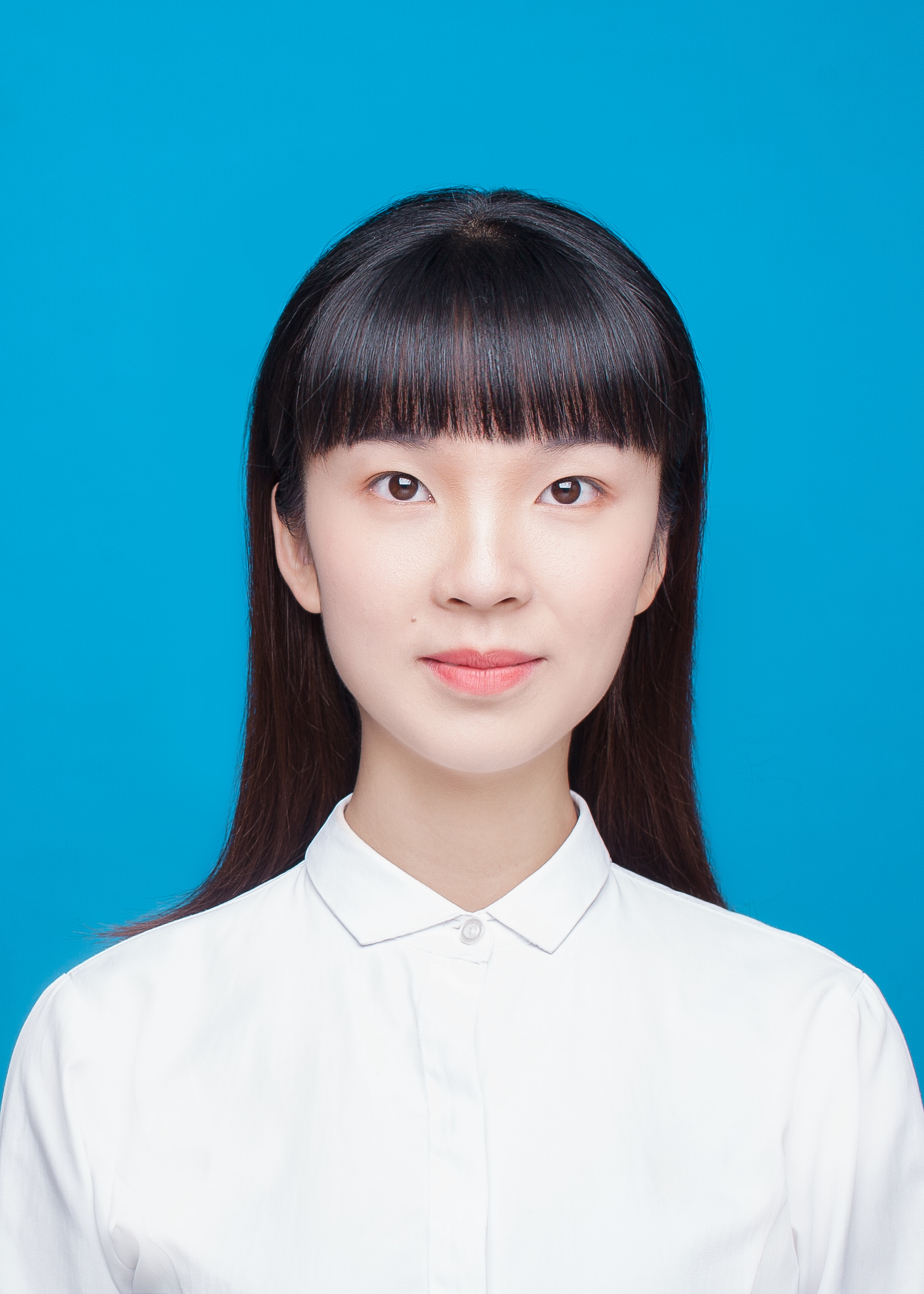}}]{Sheng Zhou} received the B.E. degree in Internet of Things from Hengyang Normal University, China, in 2020. She is currently pursuing the Ph.D. degree in the School of Computer Science and Technology, Hefei University of Technology, China. Her research interests include computer vision, natural language processing, and multimodal machine learning.
\end{IEEEbiography}
\vspace{-1cm}

\begin{IEEEbiography}[{\includegraphics[width=1in,height=1.25in, clip,keepaspectratio]{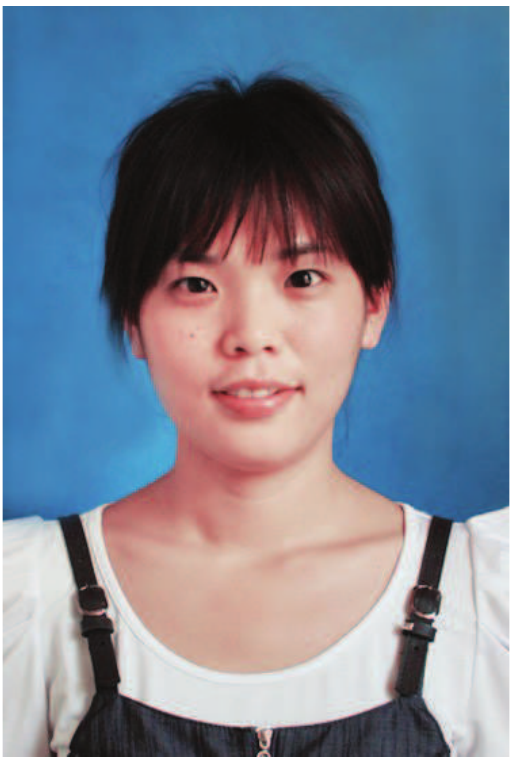}}]{Dan Guo} received the B.E. degree in computer science and technology from Yangtze University, China, in 2004, and the Ph.D. degree in System Analysis and Integration from Huazhong University of Science and Technology, China, in 2010. She is currently an Associate Professor with the School of Computer Science and Information Engineering, Hefei University of Technology, China. Her research interests include computer vision, machine learning, and intelligent multimedia content analysis.
\end{IEEEbiography}
\vspace{-1cm}

\begin{IEEEbiography}[{\includegraphics[width=1in,height=1.25in, clip,keepaspectratio]{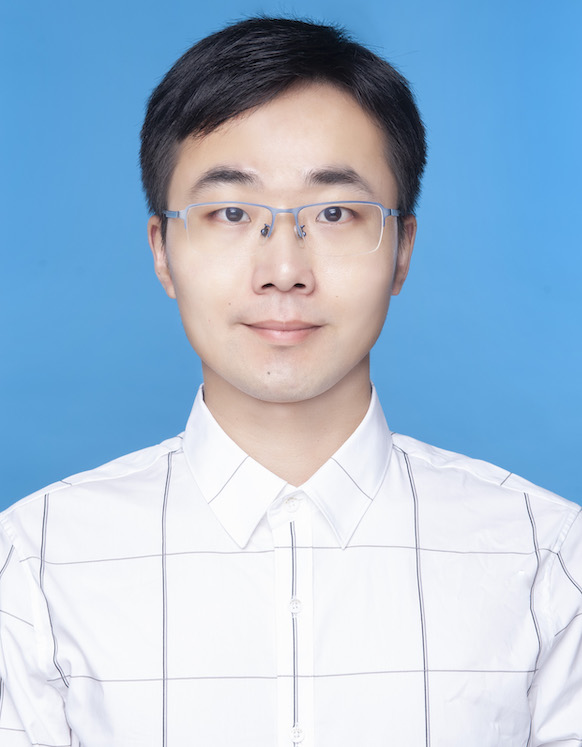}}]{Jia Li} is currently a Lecturer with Hefei University of Technology (HFUT), Hefei, China. He received his Ph.D. degree from the School of Information Science and Technology, University of Science and Technology of China (USTC), Hefei, China, in 2021. Before that, he received his B.E. degree in automation from Hefei University of Technology, China in 2016. His current research interests are in computer vision and deep learning. He has published several papers in refereed journals and conferences such as IEEE TCSVT, IEEE TITS, AAAI, and ACM MM.
\end{IEEEbiography}	
\vspace{-8cm}

\begin{IEEEbiography}[{\includegraphics[width=1in,height=1.25in, clip,keepaspectratio]{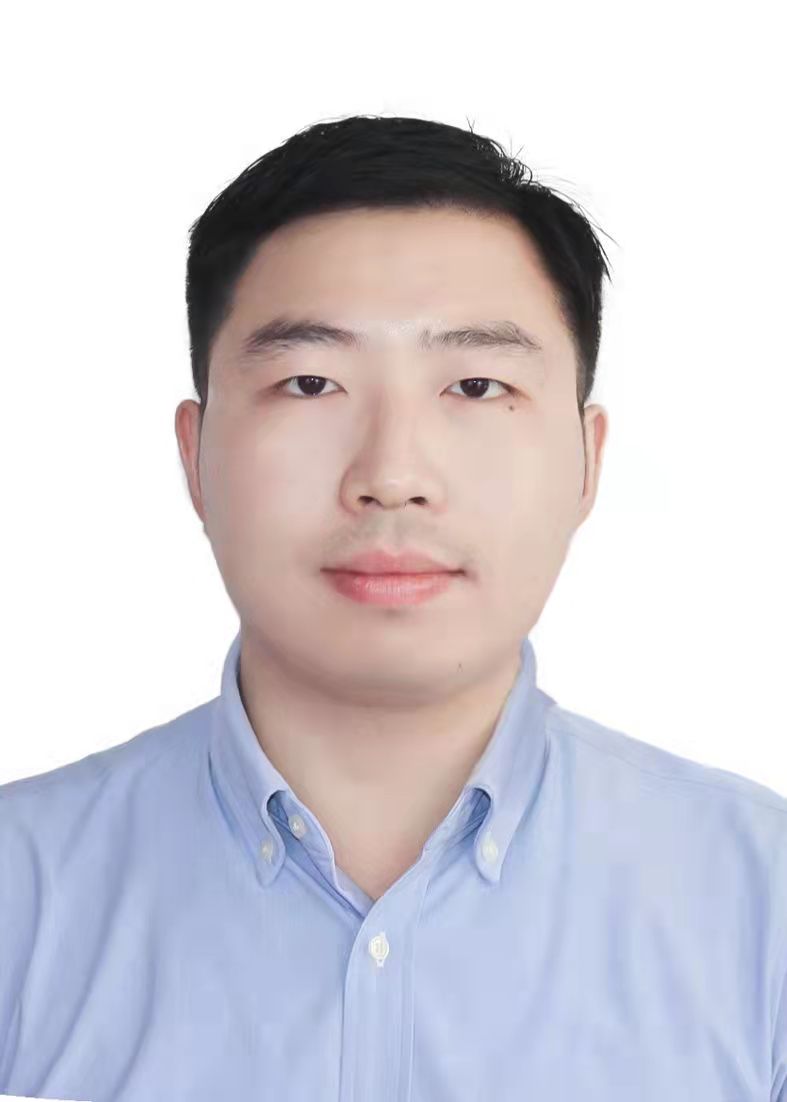}}]{Xun Yang} received his Ph.D. degree from the Hefei University of Technology, Hefei, China, in 2017.  He is currently a Professor with the Department of Electronic Engineering and Information Science, University of Science and Technology of China (USTC). From 2015 to 2017, he visited the University of Technology Sydney (UTS), Australia as a joint Ph.D student. He was a Research Fellow with the NExT++ Research Center, National University of Singapore (NUS), from 2018 to 2021. His current research interests include information retrieval, cross-media analysis and reasoning, and computer vision. He regularly serves as the PC member and the invited reviewer for top-tier conferences and prestigious journals in multimedia and artificial intelligence, like the ACM Multimedia, IJCAI, AAAI, CVPR, and ICCV. He served as the Area Chair for the ACM Multimedia 2022. He also serves as the Associate Editor for the IEEE Transactions on Big Data (TBD) journal.
\end{IEEEbiography}	
\vspace{-8cm}

\begin{IEEEbiography}[{\includegraphics[width=1in,height=1.25in, clip,keepaspectratio]{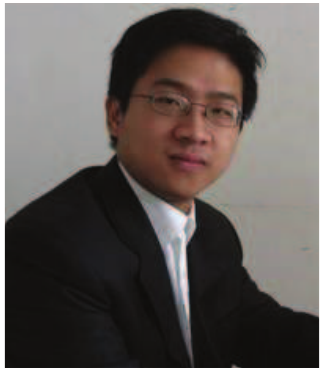}}]{Meng Wang} (SM'17) received the B.E. and Ph.D. degrees in the special class for the gifted young from the Department of Electronic Engineering and Information Science, University of Science and Technology of China, Hefei, China, in 2003 and 2008, respectively. He is currently a Professor with the Hefei University of Technology, China. He has authored over 200 book chapters, journal and conference papers in his research areas. His current research interests include multimedia content analysis, computer vision, and pattern recognition. He was a recipient of the ACM SIGMM Rising Star Award 2014. He is an Associate Editor of the IEEE TRANSACTIONS ON KNOWLEDGE AND DATA ENGINEERING, the IEEE TRANSACTIONS ON CIRCUITS AND SYSTEMS FOR VIDEO TECHNOLOGY, and the IEEE TRANSACTIONS ON NEURAL NETWORKS AND LEARNING SYSTEMS.
\end{IEEEbiography}

\end{document}